\documentclass{jfm}
\usepackage{graphicx}
\usepackage{epstopdf, epsfig}

\usepackage{amsmath}
\usepackage{nomencl}
\usepackage{subcaption}
\usepackage{cases}
\usepackage{epstopdf} 
\usepackage[retainorgcmds]{IEEEtrantools}
\usepackage{mathtools}
\usepackage{xcolor}
\usepackage[colorlinks=true,linkcolor=black,citecolor=black]{hyperref}

\newcommand{\ud}{\mathrm{d}}
\newcommand{\e}{\mathrm{e}}
\newcommand{\D}{\mathrm{D}}

\newcommand{\ri}{\mathrm{i}}
\newcommand{\rd}{\mathrm{d}}

\graphicspath{{Figs/}{IllustrationFigs/}{DataFigs/}}
\shorttitle{Temporal stability analysis of jets of lobed geometry}
\shortauthor{B. Lyu and A. P. Dowling}

\title{Temporal stability analysis of jets of lobed geometry}

\author{Benshuai Lyu\aff{1}
    \corresp{\email{bl362@cam.ac.uk}}, 
     Ann P. Dowling\aff{1}
   }

   \affiliation{\aff{1}Department of Engineering, 
University of Cambridge, Cambridge CB2 1PZ, UK}

\begin{document} 
\maketitle 
\begin{abstract}
    A 2D temporal incompressible stability analysis is carried out for lobed
    jets. The jet base flow is assumed to be parallel and of a vortex-sheet
    type. The eigenfunctions of this simplified stability problem are expanded
    using the eigenfunctions of a round jet. The original problem is then
    formulated as an innovative matrix eigenvalue problem, which can be solved
    in a very robust and efficient manner. The results show that the lobed
    geometry changes both the convection velocity and temporal growth rate of
    the instability waves. However, different modes are affected differently.
    In particular, mode $0$ is not sensitive to the geometry changes, while
    modes of higher-orders can be changed significantly. The changes become
    more pronounced as the number of lobes $N$ and the penetration ratio
    $\epsilon$ increase. Moreover, the lobed geometry can cause a previously
    degenerate eigenvalue ($\lambda_n = \lambda_{-n}$) to become non-degenerate
    ($\lambda_n \ne \lambda_{-n}$) and lead to opposite changes to the
    stability characteristics of the corresponding symmetric ($n$) and
    antisymmetric ($-n$) modes. It is also shown that each eigen-mode changes
    its shape in response to the lobes of the vortex sheet, and the degeneracy
    of an eigenvalue occurs when the vortex sheet has more symmetric planes
    than the corresponding mode shape (including both symmetric and
    antisymmetric planes). The new approach developed in this paper can be used
    to study the stability characteristics of jets of other arbitrary
    geometries in a robust and efficient manner.
\end{abstract}


\section{Introduction}
\label{sec:introduction}
Aircraft noise reduction is an urgent issue nowadays. Among the many noise
sources of an aircraft, jet noise is still a significant contributor. This is
especially true when an aircraft is taking off. While the exact role played by
jet instabilities in noise generation is still open to some debate for subsonic
jets, recent studies~\citep{Cavalieri2014, Piantanida20164350,
Lyu2016b,Lyu2016d} show that installed jet noise is dominated by the scattering
of jet instability waves by the trailing edge of aircraft wings or flaps. To
suppress or reduce installed jet noise, one can examine the possibility of
controlling these instability waves. Since instability waves are closely
related to jet mean flows, one such an approach is to modify jet mean flows to
become less axisymmetric. The feasibility of this approach, however, hinges on
the hope that the instability waves can be somehow suppressed by using a less
axisymmetric jet mean flow. This raises the question of examining the stability
characteristics of jets of general non-axisymmetric geometries.

Though jet instability is among the most heavily studied areas in fluid
mechanics~\citep{Morris2010}, the research on the characteristics of the
instability waves of non-axisymmetric jets is rather limited. Some of the early
attempts include those on elliptic and rectangular jets. These include some
analytical~\citep{Crighton1973}, numerical~\citep{Morris1988,Tam1993,Baty1995}
and some relevant experimental studies focusing on the turbulence, mixing and
acoustics of non-axisymmetric jets~\citep{Tam2000,
Li2002,Zaman2003,Miao2015,Li2001,Li2002,Hu2002}. The analytical work of
\citet{Crighton1973} showed that separable solutions can be obtained for
elliptic jets. The instability waves aligned with the major and minor axes
exhibit different behaviours. Not surprisingly, the instability waves of
rectangular jets share similar characteristics~\citep{Tam1993}. Analytical
works normally assumed the mean flow to be of a vortex sheet type to allow
mathematical derivations to proceed. For more realistic jet mean flows,
numerical methods had to be adopted. For example, \citet{Morris1988}
numerically solved the eigenvalue problem in the elliptic cylindrical
coordinates for the realistic mean flow in the initial mixing region of a jet.
It was found that all modes, no matter whether even or odd about the major
axis, have similar spatial growth rates.

Both elliptic and rectangular jets have two symmetric axes (or planes). From
the noise reduction point of view, however, it is desirable to promote
instability waves of higher-order azimuthal modes. Firstly, this is preferable
to suppress isolated jet noise. Because it has been shown that sound sources of
lower-order modes are more efficient at low frequencies for subsonic isolated
jets due to the so-called radial
compactness~\citep{Michalke1970,Mankbadi1984,Cavalieri2013}. Secondly, this is
beneficial to reduce installed jet noise. It is known that high-order
instability modes decay faster with radial distance than those of low-orders,
Hence, when they are scattered into sound by a sharp edge placed nearby, faster
decay implies weaker sound generation due to the scattering of weaker
instability waves.

Instability waves of high-order azimuthal modes may be expected to be promoted
by jet mean flows with more azimuthal periodic structures, such as lobe jets.
One of the simple lobed profiles can be described as $\sigma = a (1 + \epsilon
\cos N \theta)$, where $\sigma$ is the radius of the lobed profile, $a$ the
mean radius, $N$ the number of lobes and $\epsilon$ the penetration ratio
quantifying how large the lobes are. The open literature on the stability of
lobed jets is however very sparse~\citep{Morris2010}. In a study carried out by
\citet{Kopiev2004} on supersonic jet noise, a spatial stability analysis was
undertaken to examine the effects of weak corrugation on the instability
characteristics of a parallel vortex sheet with supersonic flow speed. A
leading-order asymptotic correction to the complex wavenumber $\alpha$ for the
round jet when $\epsilon \to 0$ was obtained. The results showed that at low
$St$, where $St$ is the Strouhal number based on the jet exit velocity $U$ and
$a$, a small $\epsilon$ may lead to a $O(\epsilon)$ change to the spatial
growth rate if the azimuthal mode number $n$ is less than the number of lobes
$N$. More interestingly, it was shown that if the sum of the two positive mode
numbers is equal to $N$, the changes to the values of their $\alpha$ have
opposite tendencies. The asymptotic solution shown in this study, however,
relies on numerically solving transcendental equations and is therefore
non-trivial to compute. Also, because the correction is restricted to
$O(\epsilon)$ or $O(\epsilon^2)$ only when $\epsilon \to 0$, it remains to be
seen to what extent the corrugation can change the stability characteristics at
a finite or large $\epsilon$. 

Of close relevance to the lobed jets are some recent studies on the stability
characteristics of chevron jets~\citep{Lajus2015,Sinha2016}. This is because
the mean flow profiles of chevrons jets are very similar to those of lobed
jets. The work of \citet{Lajus2015} was based on numerically solving the
compressible Rayleigh equation for an azimuthally periodic base flow. This
study explored the effects of azimuthal variations of the shear layer thickness
and flow radius of the mean flow on the stability characteristics. Two types of
base flow were used. The first was fitted based on a Mach $0.9$ chevron jet and
the second on a Mach $0.4$ micro-jet. For the chevron case, the results showed
that the variation of shear-layer thickness has opposite effect to that of the
radius. The combination of the two, however, results in a larger reduction of
the spatial growth rates. It was concluded that chevron is more effective in
controlling jet noise. The last section of this paper showed the effects of the
number of lobes. The preliminary results showed that the number of lobes is not
very important to the mode $0$ instability wave, which will be seen to be
consistent with the results obtained in this paper.

In the study of \citet{Sinha2016}, a viscous spatial linear stability analysis
was performed numerically using the PSE (Parabolized Stability Equation)
approach. The solutions to the parallel-flow stability equations were obtained
first to initiate the PSE. The LST (Linear Stability Theory) results showed
that the serrations reduce the spatial growth rate of the most unstable
eigen-modes of the jet, but their phase speeds are similar. For example, the
serrations appeared to reduce the spatial growth rate and increase the
convection velocity of the mode $0$ instability wave. These effects are found
to be in accord with the findings to be shown in the rest of this paper. The
PSE results were compared with the POD (Proper Orthogonal Decomposition) modes
of the near-field pressure obtained experimentally. Favourable agreement was
achieved. Similar agreement was obtained to the results from a further
investigation using an LES (Large Eddy Simulation) database. It was concluded
that the coherent hydrodynamic pressure fluctuations of jets from both round
and serrated nozzles agree reasonably to the instability modes of turbulent
mean flows.

Given the sparse analytical work on the stability of lobed jets and that the
majority of studies on this are numerically based, it is desirable to perform
some analytical studies, hoping to unveil more of the physics of lobed jets'
instability waves, such as the effects of varying $N$ and $\epsilon$ on the
instability waves of different mode numbers, and provide more insight in
understanding the jet physics. The following section performs such an analysis
within the temporal stability analysis framework, proposing an innovative
analytical method of studying how a general non-axisymmetric jet mean flow
changes the behaviour of instability waves. More importantly, the method does
not involve solving transcendental equations and would work for finite or even
large values of $\epsilon$. The method can also be used to study a wide range
of other problems in an efficient and robust manner.

\section{Temporal stability analysis for non-axisymmetric jets}
\subsection{The governing equation for non-axisymmetric vortex-sheet flows} 
Following the routine procedure of stability analysis, we decompose the flow
into base and fluctuation parts. Note that the time-average mean flow is often
taken as the base flow, and hence in this paper we use the mean flow and base
flow interchangeably. We start with the incompressible Navier-Stokes equations
since installed jet noise is relevant primarily at low Mach numbers. At this
stage, we write equations in a vector form to avoid the introduction of
coordinate systems. The momentum equation can be written as
\begin{equation}
    \frac{\D \boldsymbol{v}}{\D t} = -\frac{1}{\rho} \nabla p + \nu \nabla^2
    \boldsymbol{v},
    \label{equ:NavierStokes}
\end{equation}
where, $t$ denotes time,  $\boldsymbol{v}$ the fluid velocity, $\rho$ the flow
density, $p$ the pressure and $\nu$ the dynamic viscosity. We assume that base
flow is steady, with flow density, velocity, and pressure to be $\rho$,
$\boldsymbol{U}$ and $\bar{p}$ respectively, and that that base flow satisfies 
\begin{equation}
    \frac{\D \boldsymbol{U}}{\D t} = -\frac{1}{\rho} \nabla \bar{p} + \nu \nabla^2
    \boldsymbol{U}.
    \label{equ:BaseNavierStokes}
\end{equation}
\nomenclature[Rv]{$\boldsymbol{v}$}{Fluid velocity}
\nomenclature[Gn]{$\nu$}{Dynamic viscosity}
\nomenclature[Ru]{$\boldsymbol{U}$}{Steady base flow velocity}

The total flow field is given by the sum of the base flow and the small
perturbation. After substituting the total flow into
equation~\ref{equ:NavierStokes} and ignoring second-order quantities, we
have the following linearized equation:
\begin{equation}
    \frac{\partial \boldsymbol{v}^\prime}{\partial t} +
    \boldsymbol{U}\cdot\nabla\boldsymbol{v}^\prime + \boldsymbol{v}^\prime
    \cdot \nabla \boldsymbol{U}  = - \frac{1}{\rho} \nabla p^\prime + \nu
    \nabla^2
    \boldsymbol{v}^\prime,
    \label{equ:Linearization}
\end{equation}
where the prime symbols denote the corresponding fluctuation quantities.
%
When the Reynolds number is high, we expect that the viscous term plays a
negligible role. Hence the term $\nu \nabla^2 \boldsymbol{v}^\prime$ can be
neglected, i.e.
\begin{equation}
    \frac{\partial \boldsymbol{v}^\prime}{\partial t} +
    \boldsymbol{U}\cdot\nabla\boldsymbol{v}^\prime + \boldsymbol{v}^\prime
    \cdot \nabla \boldsymbol{U}  = - \frac{1}{\rho} \nabla p^\prime.
    \label{equ:linearisedMomentumEquationInvscid}
\end{equation}
Equation~\ref{equ:linearisedMomentumEquationInvscid}, together with the
incompressible continuity equation
\begin{equation}
    \nabla \cdot \boldsymbol{v}^\prime = 0,
    \label{equ:continuityEquationLinearized}
\end{equation}
governs the small-amplitude inviscid perturbations over a steady base flow.

\nomenclature[Rv]{$\boldsymbol{v}^\prime$}{Perturbation velocity}

The perturbation field is generally rotational for general shear base flows.
Therefore, it cannot be described using a potential function. However, for
parallel flows of a vortex-sheet type, the base flows inside and outside the
vortex sheet are both irrotational. Hence the perturbation should also be
irrotational. This suggests the existence of a potential function $\psi$ for
the velocity perturbations, i.e.
\begin{equation}
    \boldsymbol{v}^\prime = \nabla \psi,
\end{equation}
on either side of the vortex sheet. The function $\psi$ will be discontinuous
across the shear layer. From the
equation~\ref{equ:continuityEquationLinearized}, one can see that the velocity
potential satisfies the Laplace equation, i.e. 
\begin{equation}
    \nabla^2 \psi = 0.
\end{equation}
The pressure perturbation is effectively decoupled from the velocity potential
and can be easily obtained from
equation~\ref{equ:linearisedMomentumEquationInvscid}. 
\nomenclature[Gp]{$\psi$}{Potential for perturbation velocity}

The vortex-sheet simplification was used extensively in stability analysis,
due in large to the fact that an analytical dispersion relation can be
generally found~\citep{Batchelor1962,Crighton1973,Kawahara2003}. From these
dispersion relation, one can gain more insight than numerical
simulations can offer. Besides, the vortex-sheet simplification is often
permissible, particularly for analysing low-frequency instability waves.
Though a realistic jet mean flow spreads slowly and has an increasingly
thick mixing layer towards downstream, the vortex-sheet simplification should
serve as a good approximation to the realistic flow close to the jet nozzle.
Therefore, in this paper we assume the base flow to be parallel and of a
vortex-sheet type.

Since the velocity potentials exist for the vortex-sheet problem, we let
$\psi^+$ and $\psi^-$ denote the potentials outside and inside of the vortex
sheet, respectively, i.e. 
\begin{equation}
    {\boldsymbol{v}^\prime}^{\pm} = \nabla \psi^{\pm},
    \label{equ:potentialDefintion}
\end{equation}
where ${\boldsymbol{v}^\prime}^{+}$ and ${\boldsymbol{v}^\prime}^-$ denote the
velocity perturbations outside and inside the vortex sheet, respectively.
\begin{figure}
    \centering
    \includegraphics[width=0.6\textwidth]{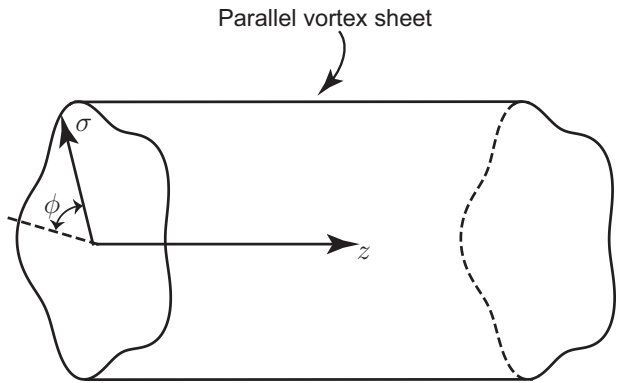}
    \caption{The cylindrical coordinate system: the $z$ axis is parallel to
	the vortex sheet, the origin is at the centre of the lobed profile and
	$\phi$ denotes the anticlockwise angle between the radial axis and the
    horizontal dashed line.}
    \label{fig:cylindricalCoordinateSystem}
\end{figure}
Considering the parallel-flow assumption, we now introduce the cylindrical
coordinates $\sigma$, $\phi$ and $z$, as shown in
figure~\ref{fig:cylindricalCoordinateSystem}. In this coordinate frame,
the velocity potentials $\psi^{\pm}(\sigma, \phi, z, t)$ satisfy the following
Laplace equation: 
\begin{equation}
    \nabla^2 \psi^{\pm}(\sigma, \phi, z, t) = 0.
    \label{equ:laplaceEquation}
\end{equation}
Note that we have not yet restricted the profiles of the vortex sheet.
It therefore can be of arbitrary geometry, such as rectangular, elliptic or
lobed. One can therefore let a general function $\mathcal{R}(\phi)$ denote the
radius of the vortex sheet at polar angle $\phi$. Consequently, the profile of
the vortex sheet can be specified as 
\begin{equation}
    \mathcal{F}(\sigma, \phi) = \sigma - \mathcal{R}(\phi)  = 0. 
    \label{equ:profile}
\end{equation}

\nomenclature[Rv]{$\boldsymbol{v}^{\prime+}$}{Perturbation velocity outside
the vortex sheet}
\nomenclature[Rv]{$\boldsymbol{v}^{\prime-}$}{Perturbation velocity inside
the vortex sheet}
\nomenclature[Gp]{$\psi^+$}{Velocity potential outside the vortex sheet}
\nomenclature[Gp]{$\psi^-$}{Velocity potential inside the vortex sheet}
\nomenclature[Rr]{$\mathcal{R}$}{Vortex sheet profile}
\nomenclature[Rf]{$\mathcal{F}$}{Function defining the vortex sheet profile}

\subsection{The eigenvalue problem}
\label{subsec:theEigenValueProblem}
\begin{figure}
    \centering
    \includegraphics[width=0.4\textwidth]{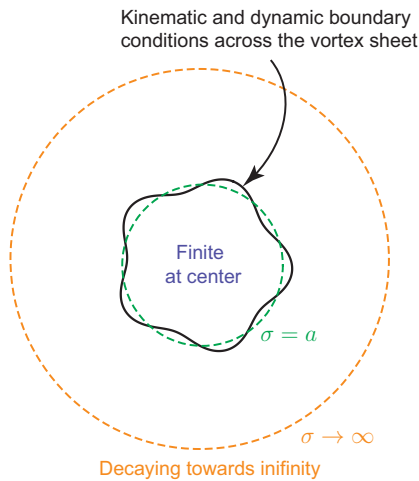}
    \caption{The schematic illustration of the boundary conditions of the
    stability problem of a parallel lobed vortex sheet.}
    \label{fig:lobedProblem}
\end{figure}

Without losing generality, one may assume
\begin{equation}
    \psi^\pm = \sum_{m=-\infty}^\infty A_m E_m^{\pm}(\sigma, \phi)
    \e^{\ri \alpha z} e^{-\ri \omega t},
    \label{equ:expansionUsingE}
\end{equation}
where $A_m$ are complex constants, $\alpha$ and $\omega$ are the streamwise
wavenumber and frequency, respectively. The functions $E_m^\pm(\sigma, \phi)$
are linearly independent of each other and each pair of them at a given $m$
satisfies both the governing equations and appropriate boundary conditions.
The functions $E_m^+(\sigma, \phi)$ and $E_m^-(\sigma, \phi)$ are therefore
defined in the two-dimensional regions outside ($\sigma > \mathcal{R}(\phi)$)
and inside ($\sigma \le \mathcal{R}(\phi)$) the vortex-sheet profile,
respectively, as shown in figure~\ref{fig:lobedProblem}.

We choose to normalize the functions $E_m^-(\sigma, \phi)$ such that for any
integer $m$
\begin{equation}
    \frac{1}{2\pi}\int_{0}^{2\pi} E_m^- (a, \phi)
    {E_m^-}^\ast(a, \phi)
    \ud \phi = 1,
    \label{equ:normalization}
\end{equation}
where $a$, as defined in section~\ref{sec:introduction} and shown in figure~\ref{fig:lobedProblem}, is
the mean radius of the vortex sheet, which is defined by
\begin{equation}
    a = \frac{1}{2\pi}\int_0^{2\pi} \mathcal{R}(\phi) \ud \phi.
\end{equation}
We cannot normalize $E_m^+(\sigma, \phi)$ according to
equation~\ref{equ:normalization}, because $E_m^-(\sigma, \phi)$ and
$E_m^+(\sigma, \phi)$ are not independent. In fact, either the dynamic or the
kinematic boundary condition is sufficient to determine $E_m^+(\sigma, \phi)$
from a given $E_m^-(\sigma, \phi)$. Therefore, $E_m^+(\sigma, \phi)$ has to be
determined such that for the same mode number $m$, the functions $E_m^+(\sigma,
\phi)$ and $E_m^-(\sigma, \phi)$ satisfy the boundary conditions. One should
note that $E_m^-(a, \phi)$ are not properly defined within some ranges of
$\phi$ (because $E_m^-(\sigma, \phi)$ is defined for $\sigma \le
    \mathcal{R}(\phi)$ and $ a > \mathcal{R}(\phi)$ in some ranges of $\phi$,
see figure~\ref{fig:lobedProblem} for example). However, they can be naturally
defined using analytical continuation, which will become clear at a later
stage. Hence, the normalization defined by equation~\ref{equ:normalization} is
valid.

\nomenclature[Ra]{$A_m$}{Expansion coefficients in front of $E_n^\pm$}
\nomenclature[Re]{$E_m^+$}{Eigenfunctions outside the vortex sheet}
\nomenclature[Re]{$E_m^-$}{Eigenfunctions inside the vortex sheet}

Substituting $E_m^\pm(\sigma, \phi)$ into equation~\ref{equ:laplaceEquation}
yields the governing equation
\begin{equation}
    \left[\frac{\partial^2}{\partial \phi^2} + \sigma\frac{\partial}{\partial
	\sigma} \big(\sigma\frac{\partial}{\partial \sigma} \big) - \alpha^2
    \sigma^2 \right] E_m^\pm(\sigma, \phi) = 0.
    \label{equ:EnEquation}
\end{equation}
Equation~\ref{equ:EnEquation} is to be solved subject to appropriate boundary
conditions. These are a finite value of $E_m^{-}(0, \phi)$, a decay behaviour
for $E_m^+(\sigma, \phi)$ as $\sigma \to \infty$, and the kinematic and dynamic
boundary conditions across the vortex sheet~\citep{Batchelor1962,Crighton1973}.
Because all the function pairs $E_m^\pm(\sigma, \phi)$ satisfy both
equation~\ref{equ:EnEquation} and their relevant boundary conditions, they are
referred to as the eigenfunctions of this system. The functions $\{E_m^\pm
(\sigma, \phi)\}$ should form a complete set of basis for the Hilbert space
determined by the equation~\ref{equ:EnEquation} and appropriate boundary
conditions. The aim of the next section is to calculate these eigenfunctions
analytically (without discretizing equation~\ref{equ:EnEquation} and then
solving it numerically).

\subsection{The calculation of eigenfunctions}
Due to the coupling of $\sigma$ and $\phi$ of the boundary conditions,
$E_m^\pm(\sigma, \phi)$ are not generally expected to be of a separable form.
However, in the case of a cylindrical vortex sheet (round jet),  the
eigenfunctions $E_m^\pm(\sigma, \phi)$ are known to be separable. Before
proceeding to the case of more general vortex sheets, it is instructive to
review the characteristics of $E_m^\pm(\sigma, \phi)$ for a cylindrical
vortex sheet.
\subsubsection{The solution for a cylindrical vortex sheet}
For the cylindrical vortex-sheet flow, the solutions were derived by
\citet{Batchelor1962} and we review them to introduce the notation, and more
importantly, these solutions form the basis for analysing more complicated
geometries in the following section. The potentials $\psi^\pm$ are known to be
able to be expanded as
\begin{equation}
    \psi^\pm = \sum_{n = -\infty}^{\infty}\Psi_n^\pm(\sigma) \e^{\ri n \phi} \e^{\ri \alpha z} \e^{-\ri \omega t}, 
\end{equation}
where the functions $\Psi_n^\pm(\sigma)$ satisfy the modified Bessel equation
\begin{equation}
    \sigma^2 \frac{\rd ^2 \Psi_n^\pm}{\rd \sigma^2} + \sigma \frac{\rd
    \Psi_n^\pm}{\rd \sigma} - \left(\alpha^2 \sigma^2  +
    n^2 \right) \Psi_n^\pm = 0.
    \label{equ:modifiedBesselEquation}
\end{equation}
Considering the boundary condition at the centre of the mean flow and at
infinity, one can show that 
\begin{equation}
    \begin{aligned}
	\Psi^{-}_n(\sigma) &= C_n^- \frac{1}{I_n(\alpha a)} I_n(\alpha \sigma), \\
	\Psi^{+}_n(\sigma) &= C_n^+ \frac{1}{K_n(\alpha a)} K_n(\alpha \sigma).
    \end{aligned}
    \label{equ:solution}
\end{equation}
where $C_n^-$ and $C_n^+$ are arbitrary complex constants and $I_n$ and $K_n$
are the modified Bessel functions of the first and second kinds, respectively.

\nomenclature[Gp]{$\Psi^+_n$}{Separated-form potentials outside the vortex
sheet}
\nomenclature[Gp]{$\Psi^-_n$}{Separated-form potentials inside the vortex
sheet}
\nomenclature[Rc]{$C_n^+$}{Expansion coefficients outside the vortex sheet}
\nomenclature[Rc]{$C_n^-$}{Expansion coefficients inside the vortex sheet}

Applying the kinematic and dynamic boundary condition on the vortex sheet, one
obtains
\begin{subnumcases}{}
    (\omega - \alpha U) \frac{\partial \Psi^{+}_n}{\partial \sigma} =
    \omega \frac{\partial \Psi^{-}_n}{\partial \sigma},
    \label{equ:twoBoundaryConditionsa}\\
    \omega \Psi^{+}_n = (\omega - \alpha U) \Psi^{-}_n,
    \label{equ:twoBoundaryConditionsb}
\end{subnumcases}
where use is made of the fact that the function set $\{\e^{\ri n \phi}\}$
is orthogonal. For a non-trivial pair of solutions to exist ($C_n^- C_n^+ \ne
0$), equations~\ref{equ:twoBoundaryConditionsa} and
\ref{equ:twoBoundaryConditionsb} can be rearranged to yield the dispersion
relation
\begin{equation}
    \Big( \frac{\alpha U}{\omega} - 1\Big)^2 = \frac{I^\prime_n(\alpha a)
    K_n(\alpha a)}{I_n(\alpha a)K_n^\prime(\alpha a)}.
    \label{equ:dispersionRelationshipCylindricalVortex}
\end{equation}

In this special case, one can verify that the eigenfunctions, $E_n^+(\sigma,
\phi)$ and $E_n^-(\sigma, \phi)$, take the separable form of
\begin{equation}
    \begin{aligned}
	&E_n^-(\sigma, \phi) = \frac{1}{I_n(\alpha a)} I_n (\alpha \sigma)
	\e^{\ri n \phi},\\
	&E_n^+(\sigma, \phi) = 
	\Big(1 - \frac{\alpha U}{\omega}\Big)
	\frac{1}{K_n(\alpha a)} K_n (\alpha \sigma) \e^{\ri n \phi}, 
    \end{aligned}
\end{equation}
where $(1 - \alpha U / \omega)$ is obtained from
equation~\ref{equ:dispersionRelationshipCylindricalVortex}. Here,
$E_n^\pm(\sigma, \phi)$ take these simple forms because the boundary condition
involves no coupling between $\sigma$ and $\phi$ (separable), and therefore
these separable solutions are just the eigenfunction of the mathematical
problem.
\subsubsection{The solution for a vortex sheet of arbitrary geometry}
For the case of a non-axisymmetric vortex sheet, the perturbations outside and inside the
vortex sheet still remain determined by the Laplace equation. However, the
boundary condition is now more complicated. If we were to find an orthogonal
coordinate system in which the vortex-sheet profile can be represented by one of the
constant coordinate axes, we may be able to find a separable solution. For
elliptic profiles, such a coordinate system exists and eigenfunctions of
separable form can be obtained~\citep{Crighton1973}. However, it seems rather
unlikely to find such a coordinate system for a general profile.

However, in light of the completeness of the orthogonal function set $\{e^{\ri
n \phi}\}$, we are still able to write the solutions inside and outside the
vortex sheet as 
\begin{equation}
    \begin{aligned}
	\psi^{-} &= \sum_{-\infty}^{\infty}C^{-}_n 
	\frac{1}{I_n(\alpha a)} I_n(\alpha \sigma)
	\e^{\ri n \phi} \e^{\ri \alpha z} \e^{-\ri \omega t}, \\
	\psi^{+} &= \sum_{-\infty}^{\infty}C^{+}_n 
	\frac{1}{K_n(\alpha a)}K_n(\alpha \sigma)
	\e^{\ri n \phi} \e^{\ri \alpha z} \e^{-\ri \omega t},
    \end{aligned}
    \label{equ:expansion2}
\end{equation}
respectively. We must emphasize here that neither the solution $I_n(\alpha
\sigma) \e^{\ri n \phi}/I_n(\alpha a)$ nor that $K_n(\alpha \sigma) \e^{\ri n
\phi} / K_n(\alpha a)$ is the eigenfunction for this problem, since they do not
satisfy the boundary conditions on the vortex sheet (although they satisfy
equation~\ref{equ:EnEquation}). However, since they form a complete set, a
suitable combination of these separable solutions which satisfies the boundary
conditions will be the eigenfunction that we aim to obtain in this section.
There may be multiple such combinations, corresponding to eigenfunctions of
different orders. For a more compact presentation of the rest of the
derivation, let
\begin{IEEEeqnarray}{rCl}
    \bar{C}_n^- &=&  C_n^- \frac{1}{I_n(\alpha a)},\\
    \bar{C}_n^+ &=& C_n^+ \frac{1}{K_n(\alpha a)}.
    \label{equ:cBarDefinition}
\end{IEEEeqnarray}
Hence, equation~\ref{equ:expansion2} becomes
\begin{equation}
    \begin{aligned}
	\psi^{-} &= \sum_{-\infty}^{\infty}\bar{C}^{-}_n 
	I_n(\alpha \sigma)
	\e^{\ri n \phi} \e^{\ri \alpha z} \e^{-\ri \omega t}, \\
	\psi^{+} &= \sum_{-\infty}^{\infty}\bar{C}^{+}_n 
	K_n(\alpha \sigma)
	\e^{\ri n \phi} \e^{\ri \alpha z} \e^{-\ri \omega t}.
    \end{aligned}
    \label{equ:expansion3}
\end{equation}

\nomenclature[Rc]{$\bar{C}^+$}{Non-normalized expansion coefficients outside
the vortex sheet}
\nomenclature[Rc]{$\bar{C}^-$}{Non-normalized expansion coefficients inside the vortex sheet}

Equation~\ref{equ:expansion3} needs to satisfy both the kinematic and dynamic
boundary conditions on the vortex sheet, i.e. when $\sigma =
\mathcal{R}(\phi)$. The two boundary conditions can be shown to be (see more
details in Appendix~\ref{app:BC}), 
\begin{subnumcases}{}
    (\omega - \alpha U) \nabla \psi^{+} \cdot \boldsymbol{n} = \omega
    \nabla \psi^{-} \cdot \boldsymbol{n} 
    \label{equ:kinematicBC},\\
    \omega \psi^{+} = (\omega - \alpha U) \psi^{-},
    \label{equ:dynamicBC}
\end{subnumcases}
where $\boldsymbol{n}$ denotes the unit vector normal to the surface of the
vortex sheet.

\section{Analysis for lobed vortex sheets}
\begin{figure}
    \begin{subfigure}[b]{0.5\textwidth}
	\centering
	\includegraphics[width=0.95\linewidth]{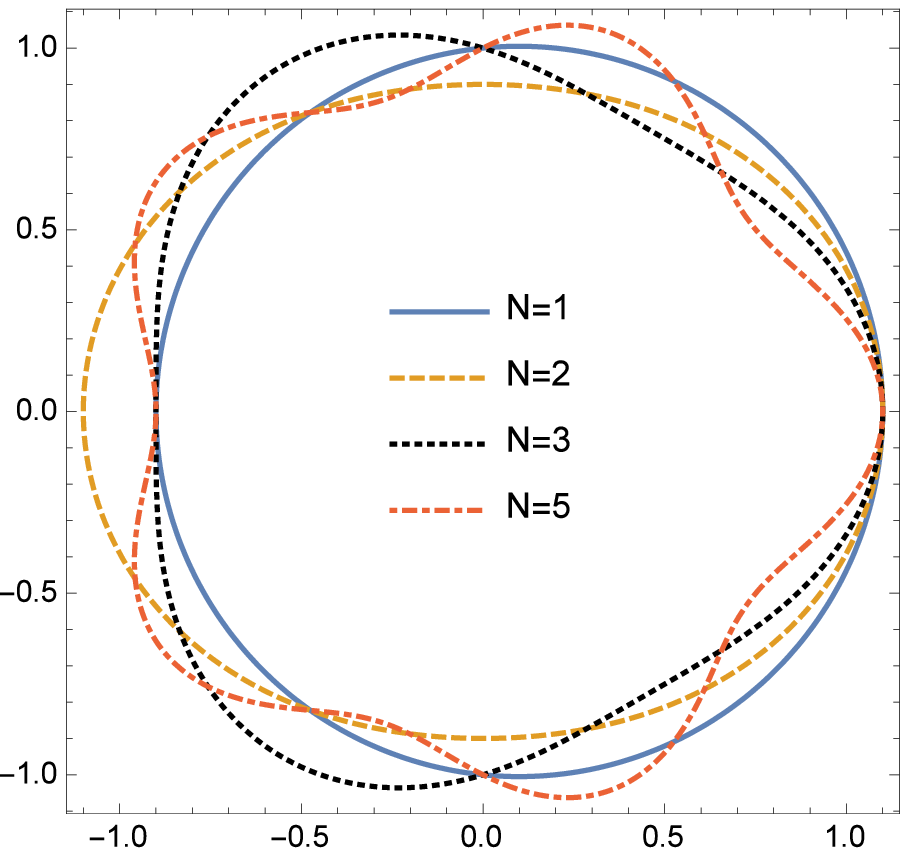}
	\caption{The weakly lobed profiles when $\epsilon = 0.1$.}
	\label{fig:weaklylobedprofile}
    \end{subfigure}
    \begin{subfigure}[b]{0.5\textwidth}
	\centering
	\includegraphics[width=0.95\linewidth]{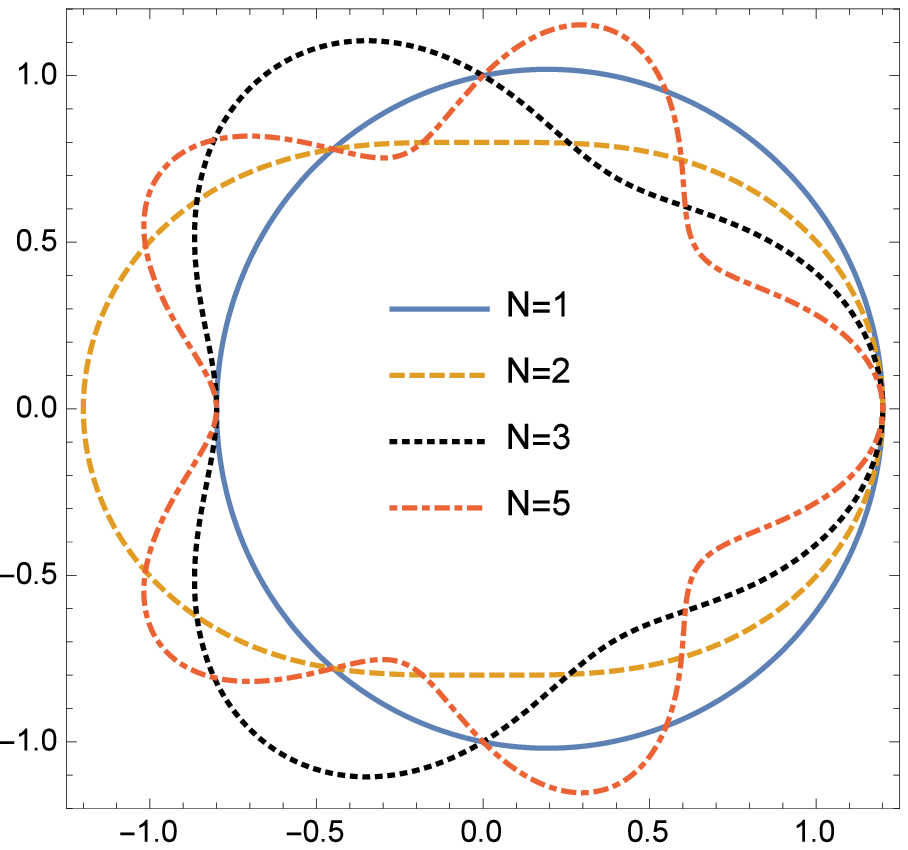}
	\caption{Strongly lobed profiles when $\epsilon = 0.2$.}
	\label{fig:stronglylobedprofiles}
    \end{subfigure}
\end{figure}

Generally, $\mathcal{R}$ can be an arbitrary function of $\phi$. Since we are
mostly concerning with the stability of lobed jets, which have a number of
identical lobes, it follows that the function $\mathcal{R}$ is a periodic
function of $\phi$. This suggests that $\mathcal{R}$ can be readily expanded
using Fourier series. As a starting point, we restrict our attention to the
simplest case mentioned in section~\ref{sec:introduction}, in which case
$\mathcal{R}$ is given by
\begin{equation}
    \mathcal{R}(\phi) = a \left( 1 + \epsilon \cos N \phi\right).
    \label{equ:functionR}
\end{equation}
For weakly lobed nozzles, $\epsilon \ll 1$, while $\epsilon \sim 0.2$
represents a strongly lobed profile. Figure~\ref{fig:weaklylobedprofile} shows
some relatively weakly lobed profiles when $\epsilon = 0.1$, while
figure~\ref{fig:stronglylobedprofiles} shows some strongly lobed profiles when
$\epsilon = 0.2$. In either case, we see that $\epsilon$ is a small quantity,
and this suggests that the Taylor expansion of a well-behaved function around
$\epsilon = 0$ should converge sufficiently quickly.  For more general lobed
profiles, function $\cos N \phi$ can be replaced by a sum over $m$ of terms with
$\phi$ dependence $\cos m N \phi$ and $\sin m N \phi$. When $\mathcal{R}$ is
given by equation~\ref{equ:functionR},  we see that 
\begin{equation}
    \boldsymbol{n} = \frac{\nabla \mathcal{F}}{\left|\nabla \mathcal{F}\right|},
    \label{equ:unitNormalVector}
\end{equation}
and
\begin{equation}
    \nabla \mathcal{F} = \boldsymbol{e}_\sigma +
    \boldsymbol{e}_{\phi}\frac{a}{\sigma} \epsilon N \sin N \phi,
    \label{equ:normalDirections}
\end{equation}
where $\boldsymbol{e}_\sigma$ and $\boldsymbol{e}_\phi$ denote the unit vectors
in the radial and azimuthal directions, respectively.
Substituting equations~\ref{equ:expansion3}, \ref{equ:functionR} and
\ref{equ:unitNormalVector} into equations~\ref{equ:kinematicBC} and
\ref{equ:dynamicBC} yields 
\begin{equation}
    \begin{dcases}
	\begin{aligned}
	    \Big(1 - \frac{\alpha U}{\omega}\Big)& \left(\sum_{n = -\infty}^\infty \bar{C}_n^+
	    K_n^\prime(\alpha a(1+\epsilon \cos N \phi)) \e^{\ri n \phi}\right.\\
	    &+ \left. 
		\frac{\epsilon N \sin N \phi}{\alpha a(1+\epsilon \cos N \phi)^2}\sum_{n = -\infty}^\infty \bar{C}_n^+
	    (\ri n) K_n(\alpha a(1+\epsilon \cos N \phi))\e^{\ri n \phi} \right) =\\ 
	    & \left(\sum_{n = -\infty}^\infty \bar{C}_n^- I_n^\prime(\alpha a(1+\epsilon \cos N \phi))
	    \e^{\ri n \phi} \right.\\
	    & + \left. \frac{\epsilon N \sin N \phi}{\alpha a(1+\epsilon \cos N \phi)^2}\sum_{n =
		-\infty}^\infty \bar{C}_n^- (\ri n) I_n(\alpha a(1+\epsilon \cos N \phi))\e^{\ri n \phi}
	    \right), 
	\end{aligned}\\
	\sum_{n = -\infty}^{\infty} \bar{C}_n^+ K_n (\alpha a(1+\epsilon \cos N \phi) ) \e^{\ri n
	\phi} = \Big(1 - \frac{\alpha U}{\omega}\Big) \sum_{n =
	-\infty}^\infty \bar{C}_n^- I_n(\alpha a(1+\epsilon \cos N \phi)) \e^{\ri n \phi}.
    \end{dcases}
    \label{equ:coupleEquationLobed}
\end{equation}

\nomenclature[Re]{$\boldsymbol{e}_\sigma$}{Unit vector in the radial direction}
\nomenclature[Re]{$\boldsymbol{e}_\phi$}{Unit vector in the azimuthal
direction}

\subsection{Weakly lobed profile}
For weakly lobed profile, $\epsilon \ll 1$, therefore we may expand both left
and right hand sides of equation~\ref{equ:coupleEquationLobed} around $\epsilon
= 0$ and keep only the first order without causing too much error. In doing so,
the first-order equations can be obtained and, after collecting the terms of
the same $\e^{\ri n \phi}$, written as
\begin{equation}
    \begin{dcases}
	\begin{aligned}
	    &\sum_{n = -\infty}^{\infty}
	    \left[\bar{C}_n^+ K_n^\prime (\alpha a)
		+ \bar{C}_{n-N}^+ \left(K_{n-N}^{\prime\prime}(\alpha a) 
		    \frac{\alpha a}{2}
		+ K_{n-N}(\alpha a) \frac{(n-N) N}{2 \alpha a}\right)\epsilon 
	    \right.\\
	    &\qquad\qquad \left. \negmedspace{}+ C_{n+N}^+ \left(K_{n+N}^{\prime\prime}(\alpha a)\frac{\alpha a}{2} -
	    K_{n+N}(\alpha a) \frac{(n+N)  N}{2 \alpha a}\right) \epsilon \right]
	    \e^{\ri n \phi} =\\ 
	    &\quad \left(\frac{1}{1 - \frac{\alpha U}{\omega}}\right) 
	    \sum_{n = -\infty}^{\infty}\left[\bar{C}_n^- I_n^\prime (\alpha a) + 
		\bar{C}_{n-N}^- \left(I_{n-N}^{\prime\prime}(\alpha a) \frac{\alpha a}{2}
	    + I_{n-N}(\alpha a) \frac{(n-N) N}{2 \alpha a}\right)\epsilon \right. \\
	    & \qquad\qquad\quad \left.\negmedspace{} + \bar{C}_{n+N}^- \left(I_{n+N}^{\prime\prime}(\alpha a)\frac{\alpha a}{2} -
	    I_{n+N}(\alpha a) \frac{(n+N) N}{2 \alpha a}\right) \epsilon \right]
	    \e^{\ri n \phi}, 
	\end{aligned}\\
	\begin{aligned}
	    &\sum_{n = -\infty}^{\infty} \left(\bar{C}_n^+ K_n (\alpha a) + 
		\bar{C}_{n-N}^+ K_{n-N}^{\prime}(\alpha a) \frac{\alpha a}{2} \epsilon +
	    \bar{C}_{n+N}^+ K_{n+N}^\prime(\alpha a)\frac{\alpha a}{2} \epsilon \right)
	    \e^{\ri n \phi} 
	    =\\
	    &\quad \Big(1 - \frac{\alpha U}{\omega}\Big) 
	    \sum_{n = -\infty}^{\infty}\left(\bar{C}_n^- I_n (\alpha a) + 
		\bar{C}_{n-N}^- I_{n-N}^{\prime}(\alpha a) \frac{\alpha a}{2} \epsilon +
	    \bar{C}_{n+N}^- I_{n+N}^\prime(\alpha a)\frac{\alpha a}{2} \epsilon \right)
	    \e^{\ri n \phi}. 
	\end{aligned}
    \end{dcases}\IEEEeqnarraynumspace
    \label{equ:coupleEquationLobedFirstOrderDifference}
\end{equation}
The above equations can be written in a more compact matrix form, from which
the effects of lobed jets can be seen more clearly. Let $\boldsymbol{K}(\alpha
a)$ denote the diagonal matrix 
\begin{equation}
    \boldsymbol{K}(\alpha a) = \textrm{diag}(
    \cdots K_{-1}(\alpha a),\, K_{0}(\alpha a),\, K_{1}(\alpha a)\cdots),
    \label{equ:matrixK}
\end{equation}
and let $\boldsymbol{K}_c(\alpha a)$ and $\boldsymbol{K}_s(\alpha a)$ to be
\begin{equation}
    \boldsymbol{K}_c(\alpha a) = 
    \begin{bmatrix*}[l]
	\hdots & \hdots \\
	\hdots K_{-N-1}(\alpha a)  & \hdots  K_{N-1}(\alpha a)\hdots\hdots\hdots \\ 
	\hdots\hdots K_{-N}(\alpha a) &\hdots\hdots K_{N}(\alpha a)\hdots\hdots\\ 
	\hdots \hdots\hdots K_{-N+1}(\alpha a) & \hdots\hdots\hdots  K_{N+1}(\alpha a) \hdots\\ 
	\hdots\hdots\hdots\hdots & \hdots\hdots\hdots\hdots\\
    \end{bmatrix*},
    \label{equ:matrixKc}
\end{equation}
and
\begin{equation}
    \boldsymbol{K}_s(\alpha a) = 
    \begin{bmatrix*}[l]
	\hdots & \hdots \\
	\hdots K_{-N-1}(\alpha a)(-N -1)  & \hdots  - K_{N-1}(\alpha a)(N-1)\hdots\hdots\hdots \\ 
	\hdots\hdots K_{-N}(\alpha a)(-N) &\hdots\hdots - K_{N}(\alpha a)N\hdots\hdots\\ 
	\hdots \hdots\hdots K_{-N+1}(\alpha a)(-N+1) & \hdots\hdots\hdots  -
	K_{N+1}(\alpha a)(N+1) \hdots\\ 
	\hdots\hdots\hdots\hdots & \hdots\hdots\hdots\hdots\\
    \end{bmatrix*}.
    \label{equ:matrixKs}
\end{equation}
Clearly, each element of the matrices $\boldsymbol{K}(\alpha a)$,
$\boldsymbol{K}_c(\alpha a)$ and $\boldsymbol{K}_s(\alpha a)$ is a function of
$\alpha a$, and consequently, these matrices are essentially function matrices
with an argument $\alpha a$. We can therefore define the $n$th derivative of
a matrix to be the matrix formed by the $n$th derivative of each element
function. For example, the first derivative of $\boldsymbol{K}_c(\alpha a)$ is
\begin{equation}
    \boldsymbol{K}_c^\prime(\alpha a) = 
    \begin{bmatrix*}[l]
	\hdots & \hdots \\
	\hdots  K_{-N-1}^\prime(\alpha a) & \hdots  K_{N-1}^\prime(\alpha a)\hdots\hdots\hdots \\ 
	\hdots\hdots K_{-N}^\prime(\alpha a) &\hdots\hdots K_{N}^\prime(\alpha a)\hdots\hdots\\ 
	\hdots \hdots\hdots K_{-N+1}^\prime(\alpha a) & \hdots\hdots\hdots K_{N+1}^\prime(\alpha a) \hdots\\ 
	\hdots\hdots\hdots\hdots & \hdots\hdots\hdots\hdots\\
    \end{bmatrix*}.
    \label{equ:matrixK2}
\end{equation}

\nomenclature[Rk]{$\boldsymbol{K}$}{Diagonal matrix}
\nomenclature[Rk]{$\boldsymbol{K}_c$}{First-order coupling matrix}
\nomenclature[Rk]{$\boldsymbol{K}_s$}{First-order coupling matrix}

If one replaces the modified Bessel function of the second kind $K_n(\alpha
a)$, in matrices $\boldsymbol{K}(\alpha a)$, $\boldsymbol{K}_c(\alpha a)$ and
$\boldsymbol{K}_s(\alpha a) $, with function $I_n(\alpha a)$, the matrices
$\boldsymbol{I}(\alpha a)$, $\boldsymbol{I}_c(\alpha a)$
and $\boldsymbol{I}_s(\alpha a)$ can be similarly defined. 
Upon defining the vector
\begin{equation}
    \boldsymbol{\bar{C}^{\pm}} = \left[ \hdots, \bar{C}_{-n}^{\pm}, \hdots,
	\bar{C}_0^{\pm}, \hdots,
    \bar{C}_n^{\pm}, \hdots \right]^T,
    \label{equ:vectorDefinition}
\end{equation}
where $[\ \,]^T$ denotes the transpose of matrix $[\ \,]$,
equation~\ref{equ:coupleEquationLobedFirstOrderDifference} can be readily
written as
\begin{equation}
    \begin{dcases}
	&\begin{aligned}
	    \Big(1 - \frac{\alpha U}{\omega}\Big)
	    &\left[\boldsymbol{K}^\prime(\alpha a) +
		\epsilon\left(\boldsymbol{K}^{\prime\prime}_{c}(\alpha a)\frac{\alpha a}{2}
		    + \boldsymbol{K}_s(\alpha a) \frac{N}{2\alpha
	    a}\right)\right]\boldsymbol{\bar{C}}^+ =\\ 
	    & \quad\quad\quad\left[\boldsymbol{I}^\prime(\alpha a) +
		\epsilon\left(\boldsymbol{I}^{\prime\prime}_{c}(\alpha a)\frac{\alpha a}{2}
		+ \boldsymbol{I}_s(\alpha a) \frac{N}{2\alpha a}\right)
	    \right]\boldsymbol{\bar{C}}^-,
	\end{aligned}\\
	&\left[\boldsymbol{K}(\alpha a) + \epsilon \boldsymbol{K}_c^\prime(\alpha
	a)\frac{\alpha a}{2}\right]\boldsymbol{\bar{C}}^+ = \Big(1 - \frac{\alpha
	U}{\omega}\Big)\left[\boldsymbol{I}(\alpha a) + \epsilon
	    \boldsymbol{I}_c^\prime(\alpha a)\frac{\alpha a}{2}
	\right]\boldsymbol{\bar{C}}^- .
    \end{dcases}
    \label{equ:matrixEquations}
\end{equation}
Equation~\ref{equ:matrixEquations} represents the dispersion relation for the
considered lobed jet to the first-order accuracy. It is worth noting that both
$\boldsymbol{K}(\alpha a)$ and $\boldsymbol{I}(\alpha a)$ are diagonal
matrices. Therefore, in the case of $\epsilon = 0$, i.e. the axisymmetric
vortex sheet, equation~\ref{equ:matrixEquations} represents a set of decoupled
dispersion relation equations. Consequently, stability analysis can be
performed for each mode individually and the results are identical to those
obtained by \citet{Batchelor1962}, as shown above. When $\epsilon \ne 0$, the
equations governing the dispersion relations are coupled equations, hence the
equations must be solved together. 

\nomenclature[Rc]{$\boldsymbol{\bar{C}}^+$}{Vector form of $\bar{C}_n^+$}
\nomenclature[Rc]{$\boldsymbol{\bar{C}}^-$}{Vector form of $\bar{C}_n^-$}

Before proceeding to solve these equations, it is informative to examine how
lobed vortex-sheet profiles affect the characteristic matrices. Firstly, one
can see that the coupling only occurs between modes $n$, $n-N$ and $n+N$. Had
we included higher order terms ($\epsilon^2, \epsilon^3\dots$), the coupling
would involve modes $n\pm kN$ ($k = 0, \pm 1, \pm 2 \dots$). This shows that
the lobed profile affects the instability waves by modulating them with its
own periodicity. Secondly, the modulating effects occur in two ways: modifying
the radial length scales and changing the normal directions of vortex sheet.
The effects of modifying the radial length scales are represented by the
$\boldsymbol{K}_c^{(i)}(\alpha a)$ and $\boldsymbol{I}_c^{(i)}(\alpha a)$
matrices ($i$ here denotes the $i$th derivative). Take the strongly-lobed
profile of $N = 2$, as shown in figure~\ref{fig:stronglylobedprofiles}, as an
example. At such a large value of $\epsilon$, the lobe profile resembles that
of an elliptic vortex sheet. Hence, the radial length scales of the major and
the minor axes are different. It is known that this causes different behaviour
for instability waves orientated with different axes~\citep{Crighton1973}. In
the dynamic boundary conditions shown in equation~\ref{equ:matrixEquations},
the $\boldsymbol{K}_c^\prime(\alpha a)$ and $\boldsymbol{I}_c^\prime(\alpha
a)$ terms account for the different length scales (to the first order
accuracy) and ensure pressure is continuous across the vortex sheet. The
$\boldsymbol{K}_s(\alpha a)$ and $\boldsymbol{I}_s(\alpha a)$ matrices, on the
other hand, account for changing of the restrictions on the normal
perturbation velocities across the vortex sheet.
Equation~\ref{equ:kinematicBC} shows that it is the normal (to the
vortex sheet) perturbation velocities that have to satisfy the jump
condition. From equation~\ref{equ:normalDirections} it is evident that the use
of lobed nozzles can significantly change the local normal directions of the
vortex sheet and hence the instability characteristics. Also, it is clear from
equation~\ref{equ:normalDirections} that the normal direction changes more
pronouncedly as $N$ increases. 


To solve equation~\ref{equ:matrixEquations}, we write the two matrix equations
in a more compact form as
\begin{subnumcases}{}
    \Big(1 - \frac{\alpha U}{\omega}\Big) \boldsymbol{K}_k
    \bar{\boldsymbol{C}}^+ = \boldsymbol{I}_k \bar{\boldsymbol{C}}^-,
    \label{equ:matrixEquationsCompact1} \\
    \boldsymbol{K}_d \bar{\boldsymbol{C}}^+ =\Big( 1 -
    \frac{\alpha U}{\omega}\Big) \boldsymbol{I}_d
    \bar{\boldsymbol{C}}^-.
    \label{equ:matrixEquationsCompact2}
\end{subnumcases}
The definitions of $\boldsymbol{K}_k$, $\boldsymbol{I}_k$, $\boldsymbol{K}_d$
and $\boldsymbol{I}_k$ should be obvious when compared with
equation~\ref{equ:matrixEquations}, and from now on we omit the argument
$\alpha a$ of relevant matrices for brevity.
Equations~\ref{equ:matrixEquationsCompact1} and
\ref{equ:matrixEquationsCompact2} are in terms of $\boldsymbol{\bar{C}^\pm}$,
it is necessary to obtain an equation in terms of $\boldsymbol{C^\pm}$
($\boldsymbol{C^\pm}$ is the column vector with elements $C_n^\pm$ rather than
$\bar{C}_n^\pm$). This is because $C_n^\pm$ are the coefficients in front of
normalized functions, and hence represent the proper amplitudes of their
corresponding eigenfunctions. On the other hand, $\bar{C}_n^\pm$ denote the
non-normalized coefficients, hence their values would depend on the amplitudes
of their eigenfunctions. For example, because the value of $I_n(\alpha a)$ at
a fixed $\alpha a$ decreases exponentially as $n$ increases, $\bar{C}_n^-$
would have to increase exponentially as $n$ increases in order to ensure a
physically meaningful result is obtained. This is clearly not suitable for
any numerical evaluations at a later stage. Therefore, it is essential to
rewrite the above two equations in terms of $C^\pm$. It is
straightforward to show $\boldsymbol{C}^+ = \boldsymbol{K}
\boldsymbol{\bar{C}}^+$ and $\boldsymbol{C}^- = \boldsymbol{I}
\boldsymbol{\bar{C}}^-$. Since both $\boldsymbol{I}$ and $\boldsymbol{K}$ are
diagonal matrices, it is trivial to calculate their inverse matrices
$\boldsymbol{I}^{-1}$ and $\boldsymbol{K}^{-1}$.
Equations~\ref{equ:matrixEquationsCompact1} and
\ref{equ:matrixEquationsCompact2} can be easily changed to 
\begin{subnumcases}{}
    \Big(1 - \frac{\alpha U}{\omega}\Big) \widetilde{\boldsymbol{K}}_k
    \boldsymbol{C}^+ = \widetilde{\boldsymbol{I}}_k \boldsymbol{C}^-,
    \label{equ:matrixEquationsNormalized1} \\
    \widetilde{\boldsymbol{K}}_d \boldsymbol{C}^+ =\Big(1 -
    \frac{\alpha U}{\omega}\Big) \widetilde{\boldsymbol{I}}_d
    \boldsymbol{C}^-,
    \label{equ:matrixEquationsNormalized2}
\end{subnumcases}
where $\widetilde{\boldsymbol{K}}_k = \boldsymbol{K}_k \boldsymbol{K}^{-1}$,
$\widetilde{\boldsymbol{K}}_d = \boldsymbol{K}_d \boldsymbol{K}^{-1}$,
$\widetilde{\boldsymbol{I}}_k = \boldsymbol{I}_k \boldsymbol{I}^{-1}$,
$\widetilde{\boldsymbol{I}}_d = \boldsymbol{I}_d \boldsymbol{I}^{-1}$. These
tilde matrices can be calculated quickly since both $\boldsymbol{I}^{-1}$ and
$\boldsymbol{K}^{-1}$ are diagonal.

\nomenclature[Rk]{$\boldsymbol{K}_k$}{Fully coupled matrix}
\nomenclature[Rk]{$\boldsymbol{K}_d$}{Fully coupled matrix}
\nomenclature[Ri]{$\boldsymbol{I}_k$}{Fully coupled matrix}
\nomenclature[Ri]{$\boldsymbol{I}_d$}{Fully coupled matrix}
\nomenclature[Rc]{$\boldsymbol{C}^+$}{Vector form of $C_n^+$}
\nomenclature[Rc]{$\boldsymbol{C}^-$}{Vector form of $C_n^-$}
\nomenclature[Rk]{$\tilde{\boldsymbol{K}}_k$}{Normalized fully coupled matrix}
\nomenclature[Rk]{$\tilde{\boldsymbol{K}}_d$}{Normalized fully coupled matrix}
\nomenclature[Ri]{$\tilde{\boldsymbol{I}}_k$}{Normalized fully coupled matrix}
\nomenclature[Ri]{$\tilde{\boldsymbol{I}}_d$}{Normalized fully coupled matrix}

From equation~\ref{equ:matrixEquationsNormalized2}, we see that
\begin{equation}
    \boldsymbol{C}^+ = \Big(1 - \frac{\alpha U}{\omega}\Big)
    \widetilde{\boldsymbol{K}}_d^{-1}
    \widetilde{\boldsymbol{I}}_d \boldsymbol{C}^{-}.
    \label{equ:secondequation}
\end{equation}
Substituting equation~\ref{equ:secondequation} into
equation~\ref{equ:matrixEquationsNormalized1}, we have
\begin{equation}
    \Big(1 - \frac{\alpha U}{\omega}\Big)^2 
    \widetilde{\boldsymbol{K}}_k \widetilde{\boldsymbol{K}}_d^{-1} 
    \widetilde{\boldsymbol{I}}_d \boldsymbol{C}^{-} =
    \widetilde{\boldsymbol{I}}_k \boldsymbol{C}^-.
    \label{equ:aftercancellation}
\end{equation}
Upon multiplying $\widetilde{\boldsymbol{I}}_k^{-1}$ on both sides of
equation~\ref{equ:aftercancellation} and defining $\boldsymbol{A} =
\widetilde{\boldsymbol{I}}_k^{-1} \widetilde{\boldsymbol{K}}_k
\widetilde{\boldsymbol{K}}_d^{-1} \widetilde{\boldsymbol{I}}_d$, we obtain the
following eigenvalue problem
\begin{equation}
    \boldsymbol{A}
    \boldsymbol{C}^- =  \lambda \boldsymbol{C}^-, 
    \label{equ:eigenvalueproblem}
\end{equation}
where 
\begin{equation}
    \lambda = \Big(1 - \frac{\alpha U}{\omega}\Big)^{-2}.
    \label{equ:eigenvalueDefinition}
\end{equation}
The matrix $\boldsymbol{A}$ is of an infinite dimension. In order to calculate
its eigenvalues in practical cases, we may drop all the modes higher than $M$
(and less than $-M$). Though we expect results to become inaccurate for large
modes close to $M$, it may yield satisfactory results for relatively low-order
modes when $M$ is taken to be adequately large. These low-order modes are of
our primary interest in this study, since high-order modes vanish sufficiently
quickly according to experimental results~\citep{Tinney2008b}. Besides, the
vortex-sheet assumption would fail for high-order modes anyway. By truncating
high-order terms, we obtain a matrix of $(2M+1) \times (2M + 1)$, and there are
$2M+1$ eigenvalues (degenerate eigenvalues are counted more than once) and
their corresponding eigenvectors. For each obtained eigenvector
$\boldsymbol{C}^-$, we can obtain the corresponding $\boldsymbol{C}^+$ easily
from equation~\ref{equ:secondequation}. The fact that the non-zero eigenvector
$\boldsymbol{C}^-$ satisfies equation~\ref{equ:eigenvalueproblem} entails that
the non-trivial velocity potential $\psi^-$, determined by $\boldsymbol{C}^-$,
and the corresponding $\psi^+$, determined by $\boldsymbol{C}^+$, satisfy both
the kinematic and dynamic boundary conditions on the vortex sheet. Therefore,
each eigenvector represents an eigenfunction of the lobed problem, i.e.
\begin{equation}
    \begin{aligned}
	&E_n^-(\sigma, \phi)  = \sum_{n = -\infty}^{\infty} C_n^-
	\frac{1}{I_n(\alpha a)}I_n(\alpha \sigma) \e^{\ri n \phi}, \\
	&E_n^+(\sigma, \phi)  = \sum_{n = -\infty}^{\infty} C_n^+
	\frac{1}{K_n(\alpha a)}K_n(\alpha \sigma) \e^{\ri n \phi}.
    \end{aligned}
    \label{equ:EnExpansion}
\end{equation}
One can readily verify that, when $\boldsymbol{C}^-$ is normalized such that
$(\boldsymbol{C}^-)^{\ast T} \boldsymbol{C}^- = 1$ and $\boldsymbol{C}^+$ is
obtained from the dynamic boundary conditions shown in
equation~\ref{equ:secondequation}, both $E_n^-(\sigma, \phi)$ and
$E_n^+(\sigma, \phi)$ are normalized as described in
Section~\ref{subsec:theEigenValueProblem}.

\nomenclature[Ra]{$\boldsymbol{A}$}{Eigen matrix}
\nomenclature[Gl]{$\lambda$}{Eigenvalue of matrix $\boldsymbol{A}$} 
\nomenclature[Rm]{$M$}{Largest mode number included}

\subsection{The mode labelling strategy}
\label{subsec:labellingStrategy}
When $\epsilon=0$, there is a well-defined mode number for each mode. For
example, mode $n$ can be defined as the eigenvector 
\begin{equation}
    \boldsymbol{C}^- = \left[ \hdots 0, \hdots, 0, \hdots,
    C_n^- = 1, \hdots \right]^T.
    \label{equ:symmetricEigenvector}
\end{equation}
When $\epsilon \ne 0$, however, the eigenvector does not posses this simple
property. Instead, the resulting eigenvector has other non-vanishing elements
besides $C_n^-$. We need to develop an unambiguous strategy to label the
eigen-modes. 

We can show that the eigenvector $\boldsymbol{C}^-$ can be always defined as
either symmetric or antisymmetric (with respect to the element of index $0$).
This is due to the rotationally symmetric property possessed by the matrix
$\boldsymbol{A}$, and for brevity we have placed detailed derivation in
Appendix~\ref{app:rotationalsymmetry}.

Based on this property, we define an eigenvector to have a mode number $n$ if
it is symmetric and 
\begin{equation}
    ||\boldsymbol{C}^-  - \boldsymbol{G}|| = \sqrt{\sum_{j=-M}^M
    (|C_j^-| - G_j)^2}
\end{equation}
yields a minimum value when $n$ varies from $-M$ to $M$ and 
\begin{equation}
    \boldsymbol{G} = \left[ \hdots, g_{-n}=\sqrt{2}/2, \hdots, 0, \hdots,
    g_n = \sqrt{2}/2, \hdots \right]^T,
\end{equation}
where $g_n$ is the element of the gauge vector $\boldsymbol{G}$. For
anti-symmetric eigenvectors we label it as $-n$ in a similar manner. For $n=0$,
it is trivial to label its mode number and it can be shown that it is a
symmetric vector. By labelling the eigenvectors in this way,
equation~\ref{equ:EnExpansion} implies that all nonnegative eigenfunctions are
even functions of $\phi$ and negative ones odd.

In the following analysis, the mode number for the obtained eigenfunction is
designated according to the above conventions. Then for each mode $n$,  we can
calculate its corresponding eigenvalue $\lambda_n$ at a given value of $\alpha
a$. The complex frequency $\omega$ can be directly obtained from $\lambda_n$
according to equation~\ref{equ:eigenvalueDefinition}. This complex number
determines both the growth rate and the convection velocity of its
corresponding instability wave. Therefore, by varying the values of $\epsilon$
and $N$, one can easily examine how different lobed geometry changes both the
growth rate and convection velocity of instability waves of different mode
numbers.

\subsection{Strongly lobed nozzle}
\label{subsec:stronglyLobedNozzle}
For strongly lobed nozzles, e.g. $\epsilon \sim 0.2$, it is necessary to
include high-order terms $\epsilon^n$. Luckily this is not a difficult
extension, and all the aforementioned procedures used to solve the eigenvalue
problem remain the same. It suffices to find the high-order coefficient
matrices and add them into equation~\ref{equ:matrixEquations}, i.e.
\begin{equation}
    \begin{aligned}
	\boldsymbol{K}_k &= \boldsymbol{K}^\prime(\alpha a) 
	+ \epsilon\Big(\boldsymbol{K}^{\prime\prime}_{c}
	    (\alpha a)\frac{\alpha a}{2}
	    + \boldsymbol{K}_s(\alpha a) \frac{N}{2\alpha
	a}\Big) + \epsilon^2 \Big(\dots\Big) + \cdots, \\
	\boldsymbol{I}_k &= \boldsymbol{I}^\prime(\alpha a) 
	+ \epsilon\Big(\boldsymbol{I}^{\prime\prime}_{c}
	    (\alpha a)\frac{\alpha a}{2}
	+ \boldsymbol{I}_s(\alpha a) \frac{N}{2\alpha a}\Big) 
	+ \epsilon^2 \Big(\dots\Big) +  \cdots,
	\\
	\boldsymbol{K}_d &= \boldsymbol{K}(\alpha a) 
	+ \epsilon \boldsymbol{K}_c^\prime(\alpha
	a)\frac{\alpha a}{2}
	+ \epsilon^2 (\dots) + \cdots,\\
	\boldsymbol{I}_d &= \boldsymbol{I}(\alpha a) + \epsilon
	\boldsymbol{I}_c^\prime(\alpha a)\frac{\alpha a}{2}
	+ \epsilon^2 (\dots) + \cdots .
    \end{aligned}
\end{equation}
It is worth noting that incorporating higher-order terms does not invalidate
the matrix $\boldsymbol{A}$ being rotationally symmetric, hence all the
previous conclusions about its eigenvectors still remain valid. Due to the
nature of higher-order modified Bessel functions, expanding them around $\alpha
a$ results in a slow convergence when $\epsilon$ is large. Therefore, the
number of high-order terms needed increases quickly as $\epsilon$ increases. It
also increases when we increase the value of $M$. However, this problem can be
overcome by expanding properly scaled modified Bessel functions. Since in this
study we only need a relatively small $M$, and the extension of incorporating
more higher-order terms can be promptly automated using computer programming,
it is not strictly necessary to expand the scaled modified Bessel functions
instead. For example, a MATLAB code has been developed that can automatically
incorporate as many orders of terms as needed. Due to its analytical nature,
the computation is very fast. For example, a comprehensive eigenvalue analysis
of order $10$ with $M = 20$ takes less than $50$ milliseconds. All the results
shown in the following sections are obtained by incorporating higher-order
terms to the order of $10$ ($\epsilon^{10}$) and with $M = 20$. A convergence
analysis, to be shown in section~\ref{subsec:convergenceAnalysis}, shows that
this is much more than necessary.

\section{Validation and Results}
As mentioned in the preceding section, both the temporal growth rate and
convection velocity of the instability waves can be readily obtained from
equation~\ref{equ:eigenvalueDefinition}. In this section, we present the rich
results obtained from this procedure.

\subsection{Convergence analysis}
\label{subsec:convergenceAnalysis}
We first examine the convergence characteristics of this new method when the
order of accuracy $m$ ($\epsilon^m$) and $M$ change. To separate the effects of
$m$ and $M$, we can fix $M$ and vary the values of $m$ and then vice versa. For
brevity, we choose to present the convection velocity and temporal growth rate
of mode $1$ instability wave and fix the number of lobes $N = 2$. 
\begin{figure}
    \centering
    \includegraphics[width=\textwidth]{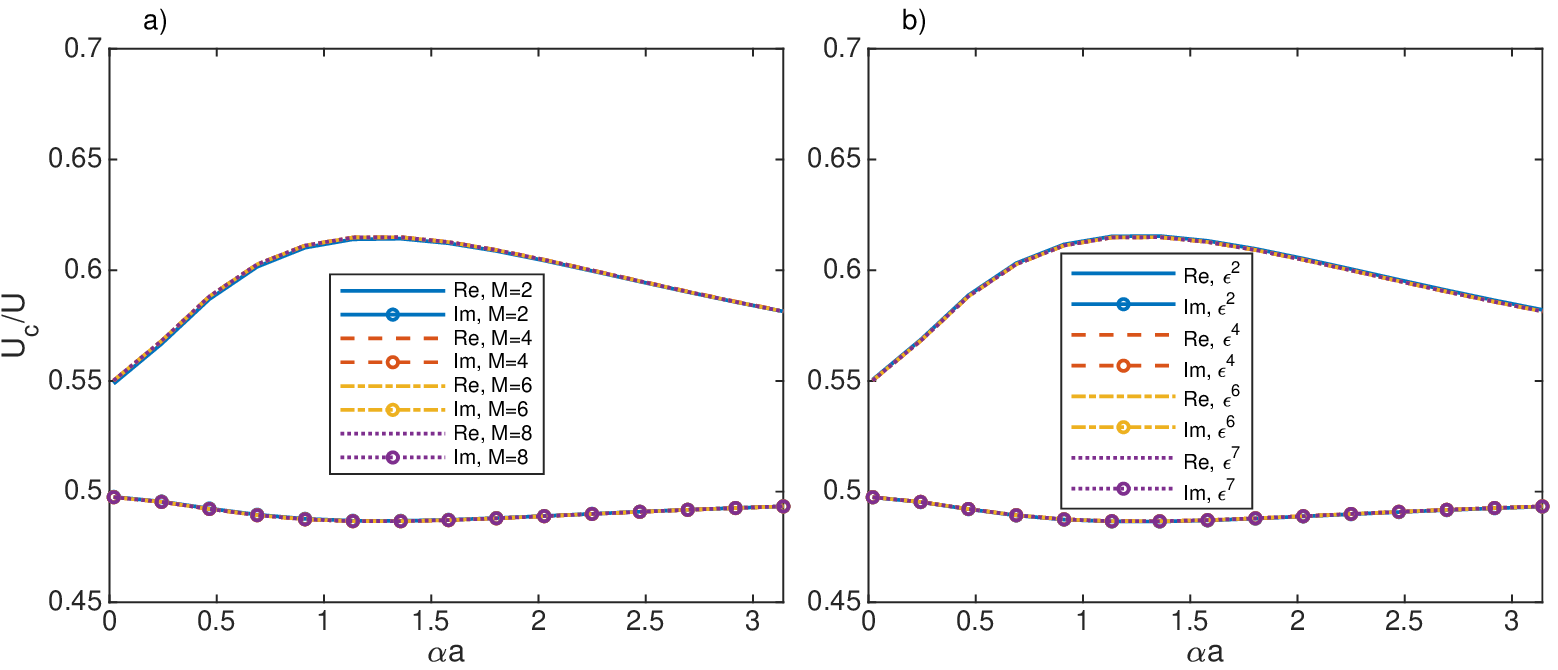}
    \caption{The convergence of the temporal growth rate and convection
velocity for the mode $1$ instability wave when $N=2$ and $\epsilon=0.1$: a)
the order of accuracy is fixed at $\epsilon^{10}$ while $M$ varies from $2$ to
$8$; b) $M$ is fixed to be $20$ while the order of accuracy varies from
$\epsilon^2$ to $\epsilon^7$.}
    \label{fig:convergenceMode1N2epsilon0p1}
\end{figure}

We first show the results for a relatively small penetration ratio, i.e.
$\epsilon = 0.1$. The results are shown in
figure~\ref{fig:convergenceMode1N2epsilon0p1}. For a compact presentation, we
define a complex number $U_c \equiv \omega/\alpha$. Therefore, the real part of
$U_c$ denotes the convection velocity while the imaginary part represents the
temporal growth rate. Figure~\ref{fig:convergenceMode1N2epsilon0p1} shows the
real and imaginary parts of $U_c$, respectively.

Figure~\ref{fig:convergenceMode1N2epsilon0p1}(a) shows the results when $m =
10$ and $M$ varies from $2$ to $8$. One can see that both the growth rates and
convection velocities are hardly distinguishable. This shows that, for a small
value of $\epsilon \approx 0.1$, $M=2$ is essentially sufficient, at least for
the mode number $1$. Of course we would need a slightly larger value of $M$ if
we were to consider higher-order modes. But this increase in $M$ is likely to
be on a small scale, because we are only interested in low-order modes
as only those are physically relevant in experiments.
Figure~\ref{fig:convergenceMode1N2epsilon0p1}(b) shows the results when $M$ is
fixed at $20$ and $m$ varies from $2$ to $7$. Similar to
figure~\ref{fig:convergenceMode1N2epsilon0p1}(a), the lines are nearly on top
of each other. This shows that for $\epsilon \approx 0.1$, a second-order
accuracy is sufficient for a good convergence.

\begin{figure}
    \centering
    \includegraphics[width=\textwidth]{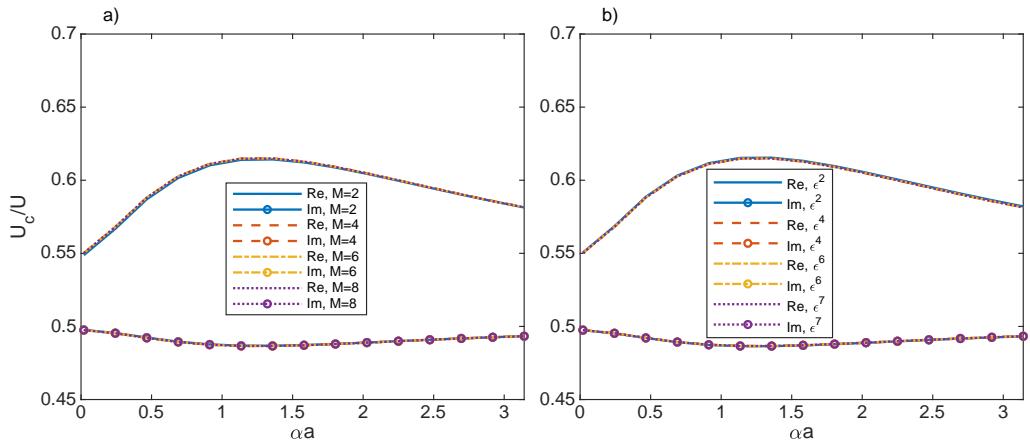}
    \caption{The convergence of the temporal growth rate and convection
velocity for the mode $1$ instability wave when $N=2$ and $\epsilon=0.2$: a)
the order of accuracy is fixed at $\epsilon^{10}$ while $M$ varies from $2$ to
$8$; b) $M$ is fixed to be $20$ while the order of accuracy varies from
$\epsilon^2$ to $\epsilon^7$.}
    \label{fig:convergenceMode1N2epsilon0p2}
\end{figure}

As mentioned in section~\ref{subsec:stronglyLobedNozzle}, due to the nature of
the modified Bessel functions, the number of high-order $\epsilon^n$ terms
needed increases quickly as $\epsilon$ increases. To show that $M = 20$ and $m
= 10$ is also sufficient for a large value of $\epsilon$, we present the
convergence characteristics of this analysis when $\epsilon = 0.2$ in
figure~\ref{fig:convergenceMode1N2epsilon0p2}.

Figure~\ref{fig:convergenceMode1N2epsilon0p2}(a) shows the results when $m =
10$ and $M$ varies from $2$ to $8$. We can see that there is an observable
difference between the convection velocities calculated using $M=2$ and $M=4$.
However, there is little change between the results for $M = 4$ and $M = 8$.
This shows that for strongly lobed geometry at $\epsilon \approx 0.2$, $M$ must
be at least $4$. On the other hand, it is interesting to see that the temporal
growth rate is much less sensitive to the change in $M$ than the convection
velocity. Figure~\ref{fig:convergenceMode1N2epsilon0p2}(b) similarly shows the
results for a fixed number $M = 20$ and a varying $m$. The observation is
similar to figure~\ref{fig:convergenceMode1N2epsilon0p2}(a), and it shows that
for $\epsilon \approx 0.2$, a high order accuracy up to $\epsilon^4$ or
$\epsilon^6$ is recommended. In summary, the results in this section show that
$M=20$ and $n =10$ can ensure the obtained eigen-solutions are well converged.

\subsection{Validation}
\label{subsubsec:validation}
\begin{figure}
    \centering
    \includegraphics[width=\textwidth]{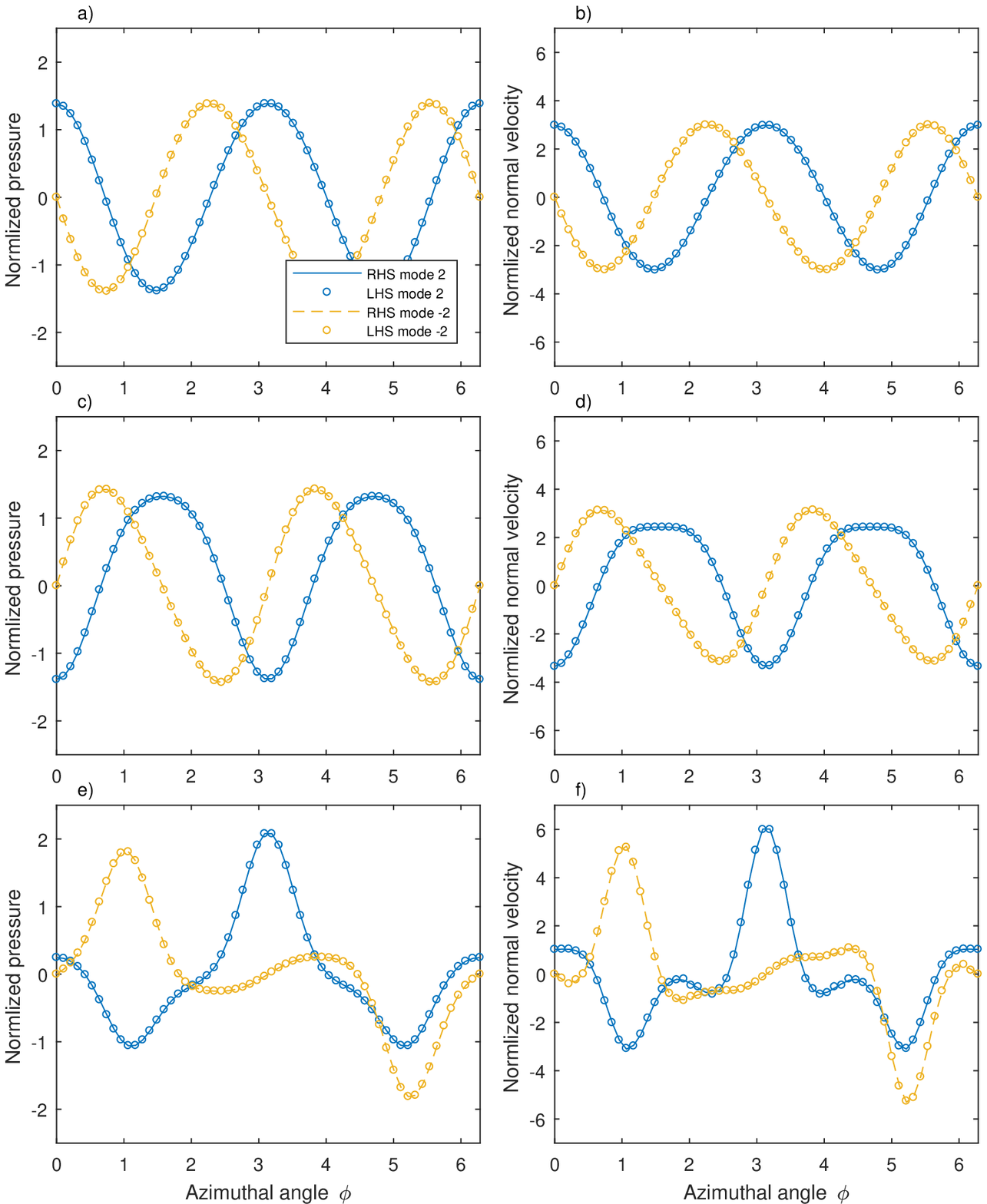}
    \caption{The match validation of the dynamic (left) and kinematic (right)
    boundary conditions for the eigenfunctions of modes $\pm 2$ when the number
    of lobes $N$ are different: a, b) $N = 1$; c, d) $N = 2$; e, f) $N = 3$.
    The normalized wavenumber is $\alpha a = 1$ and the lobed profile has a
penetration ratio $\epsilon = 0.1$.}
\label{fig:validateMode2BCsVaryingN}
\end{figure}

Section~\ref{subsec:convergenceAnalysis} merely shows that $M=20$ and $n=10$
are sufficient for a good convergence. A converged solution, however, does not
always imply a correct one. Consequently, before presenting any results, it is
necessary to validate this new analysis framework. Luckily, it is very
straightforward to do so. Since the entire analysis is devoted to calculating
the eigenfunctions that satisfy both the kinematic and dynamic boundary
conditions on the vortex sheet, we can examine the obtained eigenfunctions to
ensure that they do indeed satisfy the two boundary conditions. More precisely,
we can show that for each pair of eigenfunctions obtained above, we have, on
the vortex sheet,
\begin{subnumcases}{}
    (1- \alpha U /\omega)\nabla 
    E_m^{+}(\sigma, \phi) \cdot \boldsymbol{n} = 
    \nabla E_m^{-}(\sigma, \phi)\cdot \boldsymbol{n},
    \label{equ:kinematicBCUsingE}\\
    1 / (1 - \alpha U / \omega) E_m^{+}(\sigma, \phi) = 
    E_m^{-}(\sigma, \phi).\label{equ:dynamicBCUsingE}
\end{subnumcases}
The obtained eigen-solution will automatically satisfy the Laplace equation
both inside and outside the vortex sheet, because we have chosen to expand the
eigen-solution using a set of basis functions which are already the solutions
to the Laplace equation. Therefore, it is sufficient to validate the method by
only verifying that the boundary conditions are met. In the rest of this paper,
we refer to both sides of equation~\ref{equ:kinematicBCUsingE} and
\ref{equ:dynamicBCUsingE} as the normalized normal perturbation velocity and
pressure, respectively. One can first evaluate both sides of the above two
equations on the vortex sheet (similar to
equation~\ref{equ:coupleEquationLobed}) and plot them together. If the results
are accurate, the normalized normal perturbation velocity and pressure would
collapse. Since the eigenvector of mode $n$ ($n \ge 0$) is symmetric, if it is
also real, then it is effortless to show that the imaginary parts of
$E_n^-(\sigma, \phi)$ are strictly zero. Similarly, the real parts of
$E_{-n}^-(\sigma, \phi)$ also vanish when the eigenvector of mode $-n$ is real.
For the results shown in this paper, all the eigenvectors are real. Therefore,
for mode $n$ ($n \ge 0$), it suffices to plot only the real parts of
equations~\ref{equ:kinematicBCUsingE} and \ref{equ:dynamicBCUsingE}. Likewise,
for mode $-n$ ($n>0$), one only needs to plot the imaginary parts. For the sake
of brevity, we only show the results for modes $-2$ and $2$. All other modes of
interest have similar level of agreement. The results are shown in
figure~\ref{fig:validateMode2BCsVaryingN}. 

Figure~\ref{fig:validateMode2BCsVaryingN} shows how the boundary conditions are
satisfied by the eigenfunctions of modes $\pm 2$. These results are obtained
with $\alpha a = 1$ and $\epsilon  = 0.1$.
Figures~\ref{fig:validateMode2BCsVaryingN}(a) and
\ref{fig:validateMode2BCsVaryingN}(b) show the matches of the left and right
sides of equations~\ref{equ:dynamicBCUsingE} and \ref{equ:kinematicBCUsingE},
respectively, when the number of lobes is $N = 1$. From
figure~\ref{fig:weaklylobedprofile}, one can see that when $N = 1$, the lobed
profile is not very significantly different from the axisymmetric one. The
single lobe merely causes a displacement of the profile centre of the circular
vortex sheet. Therefore, one expects that the dynamics of instability waves
shall remain largely unchanged. Figure~\ref{fig:validateMode2BCsVaryingN}(a,b)
indeed confirms this. Modes $2$ and $-2$ resemble those of an axisymmetric
vortex sheet in every aspect.  When $N$ increases to $2$, the effects of lobes
are more pronounced, as shown in
figure~\ref{fig:validateMode2BCsVaryingN}(c,d). In particular, the symmetric
(with respect to $\phi = 0)$ eigenfunction shown in
figure~\ref{fig:validateMode2BCsVaryingN}(d) starts to deviate from the normal
$-\cos2\phi$ function, with its peak somewhat flattened. The mode $-2$, on the
other hand, remains largely similar to $\sin2\phi$. When $N$ increases to $3$,
both modes $2$ and $-2$ respond to the change of vortex-sheet geometry and
change their shapes significantly. 

It is, however, worth noting that, for $N = 1$ and $N = 2$, the eigenvalues
$\lambda_2 \ne \lambda_{-2}$. Therefore, both $E_2^-(\sigma, \phi)$ and
$E_{-2}^-(\sigma, \phi)$ have a standing wave pattern with respect to $\phi$
and they cannot be combined to produce a travelling wave. However, for $N = 3$
we find that $\lambda_2 = \lambda_{-2}$. Therefore, a travelling-wave pattern
can be obtained. The important observation is that, for all the six figures,
exceptionally good matches of the pressure and normal velocity across the layer
of vortex sheet are achieved. The excellent agreement seen from figure~
\ref{fig:validateMode2BCsVaryingN} shows that the analytical framework
developed in this section works exceptionally well, at least for the low-order
modes like those we have just shown. 
%
\subsection{The effects of lobed profiles on the convection velocity and
growth rate of instability waves}
\label{subsec:effects}
Having validated the analytical framework, we are now in a position to examine
the effects of lobed profiles on the convection velocity and growth rate of
instability waves. In the rest of this section, we plot both quantities versus
the normalized frequency $\alpha a$, for lobed profiles of different geometry.

\begin{figure}
    \centering
    \includegraphics[width=\textwidth]{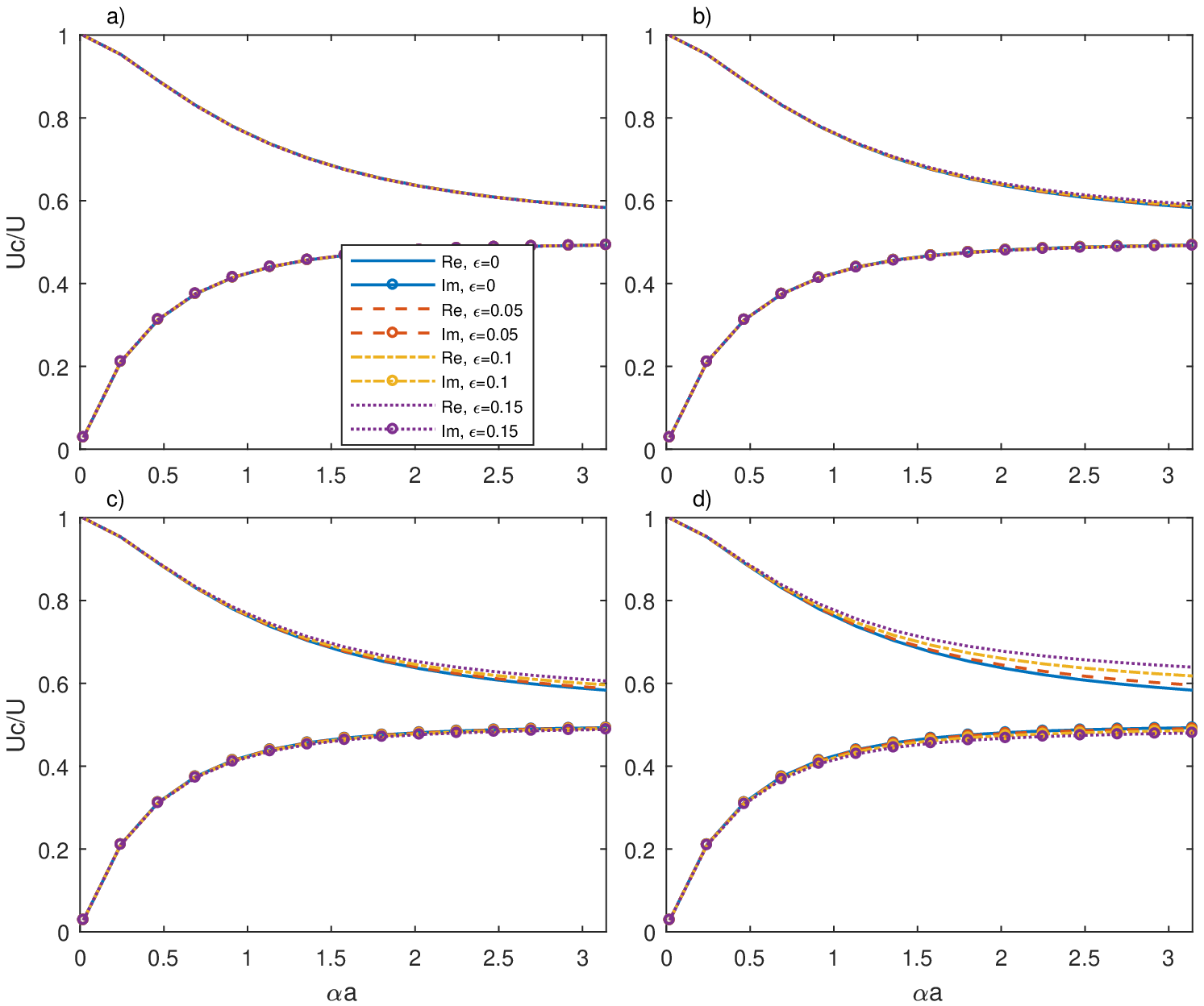}
    \caption{The convection velocity and growth rate of the mode $0$th jet
    instability waves for vortex sheets of different lobed geometry: a) $N =
1$; b) $N = 2$; c) $N = 3$; d) $N = 5$.}
\label{fig:UcVsAlphaaMode0}
\end{figure}

We start from showing results for mode $0$. These are shown in
figure~\ref{fig:UcVsAlphaaMode0}. Figure~\ref{fig:UcVsAlphaaMode0}(a) shows the
convection velocity and growth rate for a single-lobe profile. To facilitate a
direct comparison to the results of a cylindrical vortex sheet, both quantities
are plotted when $\epsilon = 0$ first. As can be seen, when the frequency
$\alpha a$ increases, the normalized convection velocity ($\Re{U_c/U}$)
decreases from unity to around $0.5$, whereas the normalized growth rate
($\Im{U_c/U}$) increases from $0$ to the same limit value. However, it shows
that a single lobe does not cause any observable changes to the characteristics
of mode $0$ at all frequencies, no matter what value of $\epsilon$ is used. The
same conclusion can be reached for $N = 2$, which is shown in
figure~\ref{fig:UcVsAlphaaMode0}(b). Increasing the number of lobes to
$3$, however, starts to cause a slightly larger convection velocity and a
marginally lower growth rate. These results are shown in
figure~\ref{fig:UcVsAlphaaMode0}(c), from which we see that the changes are
only observable at high frequencies. However, they become more pronounced when
the penetration ratio $\epsilon$ increases. The increase of the convection
velocity and the reduction of the temporal growth rate, caused by lobed
vortex-sheet profiles, are more evident when the number of lobes is increased
to $5$. As shown in figure~\ref{fig:UcVsAlphaaMode0}(d), at high frequencies
($\alpha a > 1.5$), a large penetration ratio results in an effective rise of
the convection velocity and a less effective drop of the growth rate. However,
one should note that although these changes are observable, they are not in any
way significant.

The fact that serrations increase the convection velocity and reduce growth
rates of mode $0$ is in accord with the findings of \citet{Lajus2015} and
\citet{Sinha2016}. However, it should be noted that in the work of
\citet{Sinha2016}, there also exists a low-frequency band where the spatial
convection velocity is slightly reduced. Such a difference might be caused by
the difference between temporal and spatial analysis, the difference in jet
mean flow profiles or the difference in the shear layer thickness. 

\begin{figure}
    \centering
    \includegraphics[width=\textwidth]{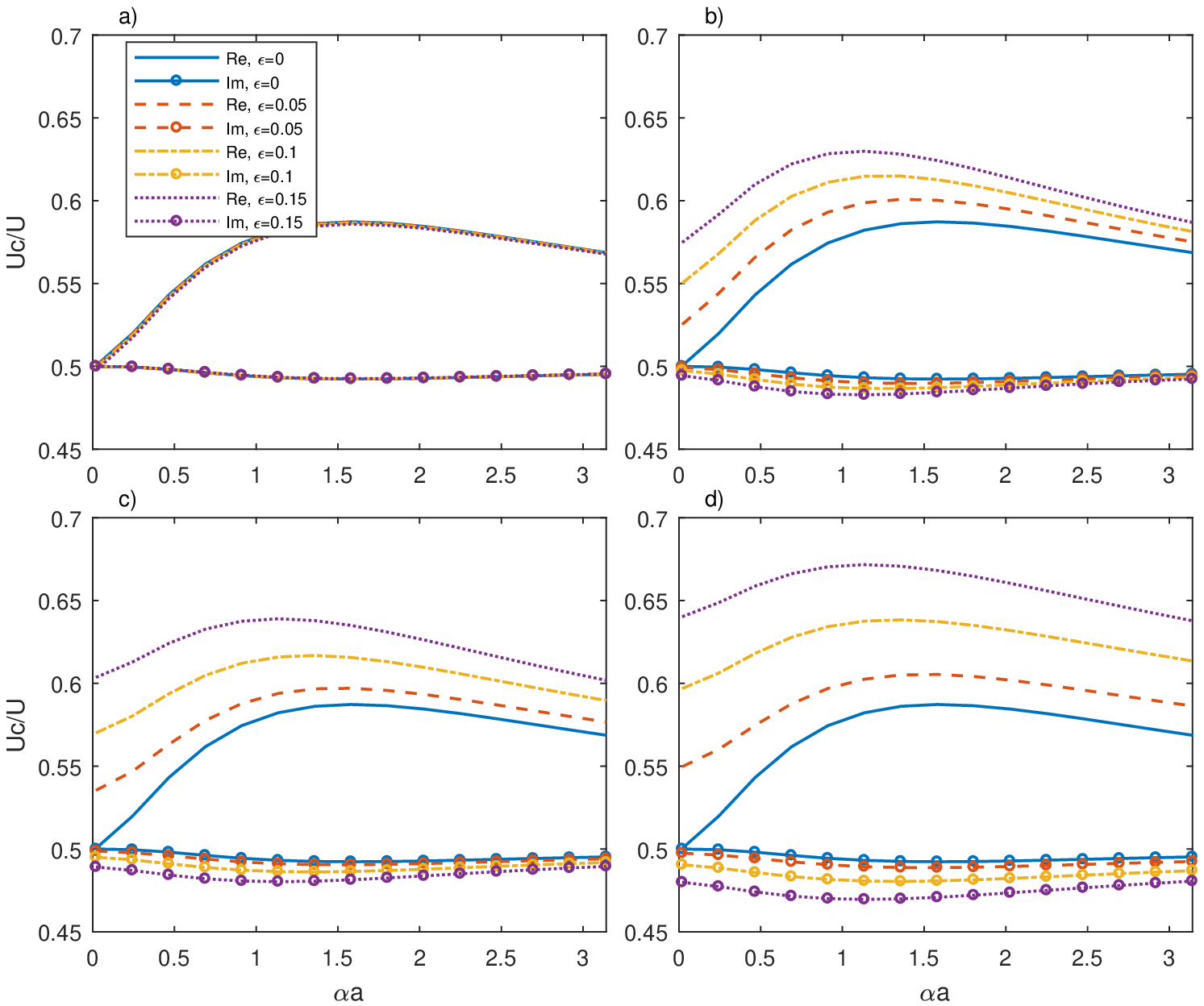}
    \caption{The convection velocity and growth rate of the mode $1$ jet
    instability waves for vortex sheets of different lobed geometry: a) $N =
1$; b) $N = 2$; c) $N = 3$ d) $N = 5$.}
\label{fig:UcVsAlphaaMode1}
\end{figure}

Figure~\ref{fig:UcVsAlphaaMode0} shows that instability waves of mode $0$ are
not very sensitive to either the number of lobes or the penetration ratio.
Figure~\ref{fig:UcVsAlphaaMode1}, however, shows a different story for mode
$1$. The eigenfunctions corresponding to mode $1$ are even functions of $\phi$.
Figure~\ref{fig:UcVsAlphaaMode1}(a) still indicates that a single lobe does not
noticeably change the characteristics of the mode $1$ instability waves.
This is somewhat expected. Because, as we observed before, one single lobe
merely causes a displacement of the profile's geometrical centre. Therefore,
the physics should more or less stay the same as that of an axisymmetric
vortex-sheet. This is consistent with the results shown in
figure~\ref{fig:UcVsAlphaaMode1}(a).
However, figure~\ref{fig:UcVsAlphaaMode1}(b) shows that the use of $2$ lobes
leads to a pronounced increase of the convection velocity, and a slight
decrease of the temporal growth rate. In contrast to those shown in
figure~\ref{fig:UcVsAlphaaMode0}, the increase of the convection velocity is
more marked at low frequencies. Similarly, increasing $\epsilon$ results in a
stronger rise of the convection velocity. It is very interesting o note that
the rise of the convection velocity is nearly linear with respect to
$\epsilon$. The decrease of the temporal growth rate, however, is most notable
in the intermediate frequency range, and the maximum decrease is very small.
Figure~\ref{fig:UcVsAlphaaMode1}(c) shows the results for $N = 3$. Compared to
figure~\ref{fig:UcVsAlphaaMode1}(b), the effects of lobes on both the
convection velocity and the growth rate appear to be more effective. The most
pronounced change, however, occurs when $N$ increases to $5$, as shown in
figure~\ref{fig:UcVsAlphaaMode1}(d). Again, these results are consistent with
the findings of \citet{Sinha2016}. The dependence of both the convection
velocity and the temporal growth rate on $\epsilon$ appears to linear. But the
relative change of the former is significantly larger than that of the latter.

\begin{figure}
    \centering
    \includegraphics[width=\textwidth]{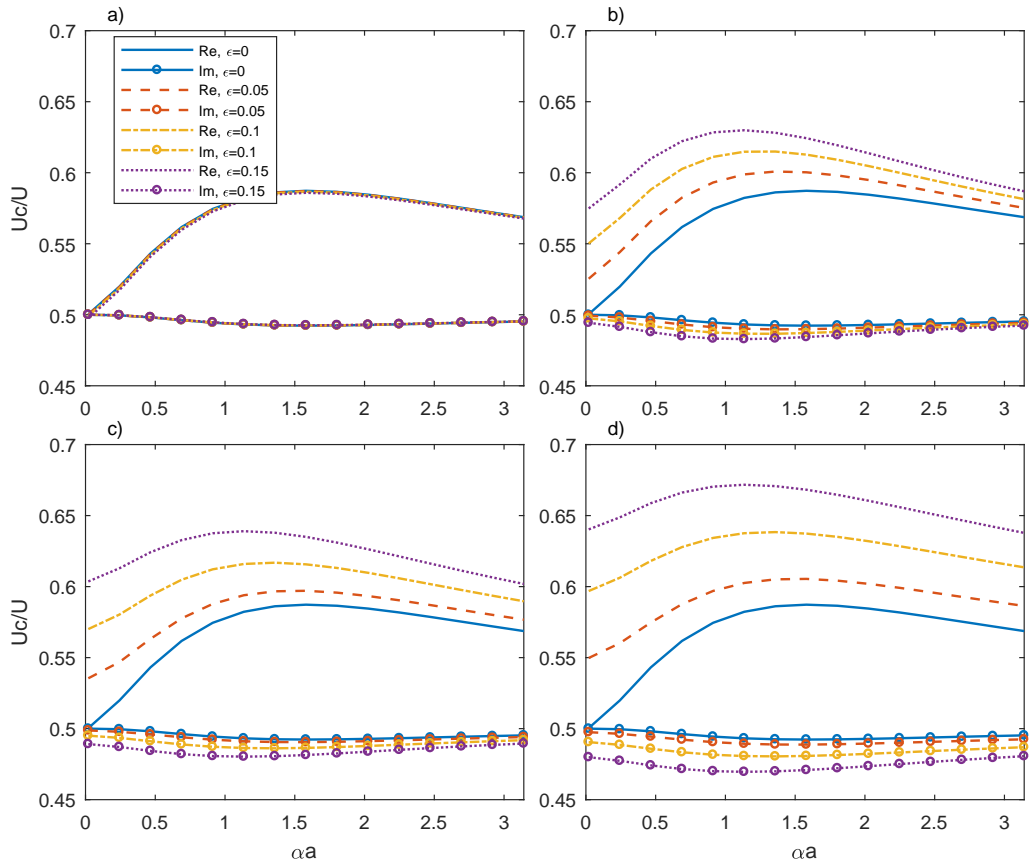}
    \caption{The convection velocity and growth rate of the mode $-1$ jet
    instability waves for vortex sheets of different lobed geometry: a) $N =
1$; b) $N = 2$; c) $N = 3$; d) $N = 5$.}
\label{fig:UcVsAlphaaMode-1}
\end{figure}

Figure~\ref{fig:UcVsAlphaaMode-1} presents the results for the mode $-1$. The
eigenfunctions are odd functions of $\phi$. Still,
figure~\ref{fig:UcVsAlphaaMode-1}(a) does not show observable changes when
$\epsilon$ increases. However, zooming in this figure, one can see that the
convection velocity is weakly increasing as $\epsilon$ increases, and this is
opposite to that observed in figure~\ref{fig:UcVsAlphaaMode1}(a)! This
different behaviour is because $\lambda_1$ is not identical to $\lambda_{-1}$
any more. Therefore, the odd and even eigenfunctions are now independent of
each other and they change in opposite ways as $\epsilon$
increases. If figures~\ref{fig:UcVsAlphaaMode-1}(a) and
\ref{fig:UcVsAlphaaMode1}(a) are too similar to each other to make this trend
clear, figure~\ref{fig:UcVsAlphaaMode-1}(b) makes it much more evident.
The number of lobes is now $2$, and the lobed profile is approximately
elliptic. Instead of obtaining higher convection velocities, increasing
$\epsilon$ from $0$ now results in increasingly smaller convection velocities.
The temporal growth rate, on the other hand, starts to drop at low frequencies
($\alpha a < 0.5$), but gradually changes to increase at high frequencies,
although both are on a small scale. The different eigenvalues of modes $\pm 1$,
hence distinctive characteristics of instability waves of modes $\pm 1$, are
consistent with the analytical results obtained by \citet{Crighton1973} for
elliptic vortex sheets and also agree with the findings in the spatial
stability analysis carried out by \citet{Kopiev2004} (because the mode number
$1$ is a half of the number of lobes $N=2$).
Figure~\ref{fig:UcVsAlphaaMode-1}(c,d) shows the results for $N = 3$ and $N =
5$, respectively. One can easily verify that they are identical to those shown
in figure~\ref{fig:UcVsAlphaaMode1}(c, d). This is because, for both $N = 3$
and $N = 5$, the eigenvalues $\lambda_n$ remain identical to $\lambda_{-n}$. As
discussed in Section~\ref{subsubsec:validation}, this also implies that
azimuthally travelling waves can exist, in contrast to the case of $\lambda_n
\ne \lambda_{-n}$, where only azimuthally standing waves are allowed.

\begin{figure}
    \centering
    \includegraphics[width=\textwidth]{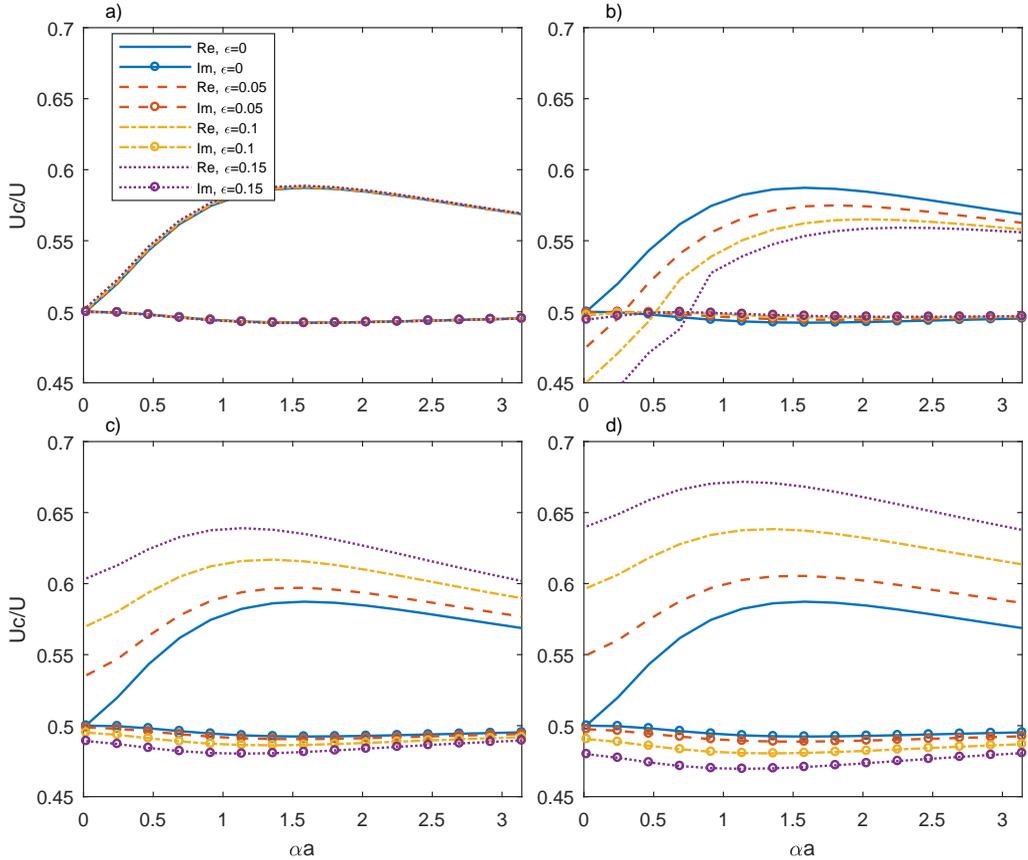}
    \caption{The convection velocity and growth rate of the mode $2$ jet
    instability waves for vortex sheets of different lobed geometry: a) $N =
1$; b) $N = 2$; c) $N = 3$; d) $N = 5$.}
\label{fig:UcVsAlphaaMode2}
\end{figure}

\begin{figure}
    \centering
    \includegraphics[width=\textwidth]{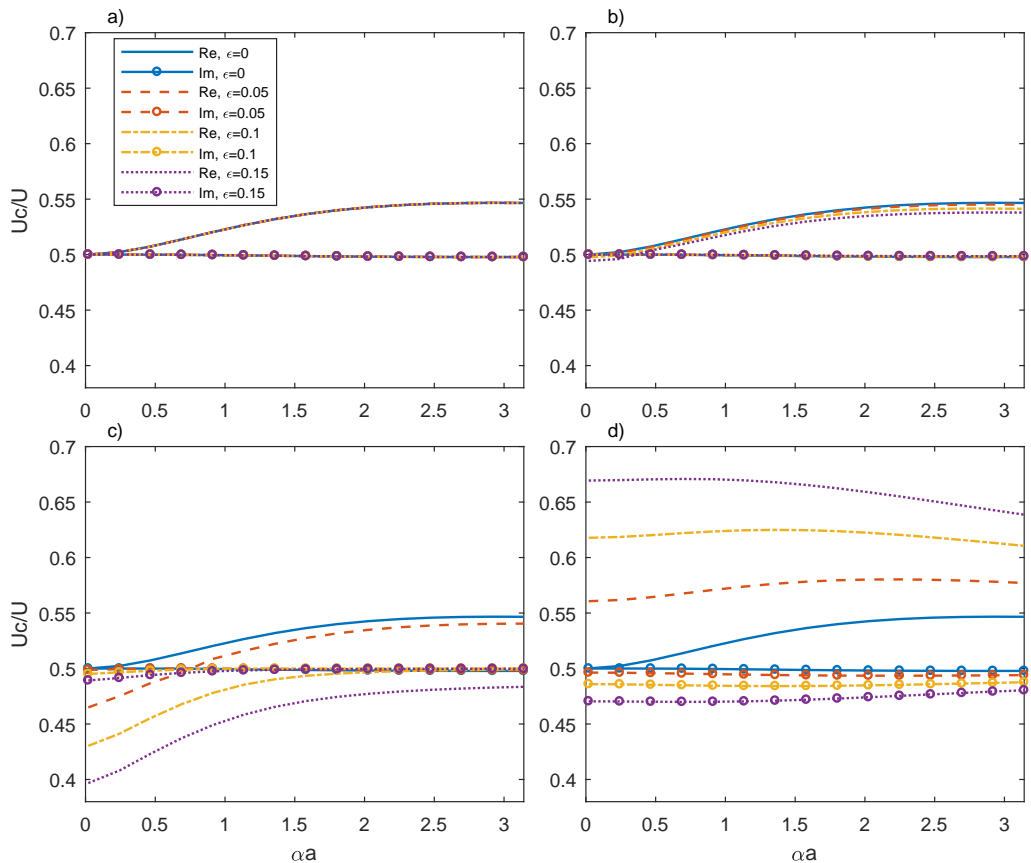}
    \caption{The convection velocity and growth rate of the mode $-2$ jet
    instability waves for vortex sheets of different lobed geometry: a) $N =
1$; b) $N = 2$; c) $N = 3$; d) $N = 5$.}
\label{fig:UcVsAlphaaMode-2}
\end{figure}

Figure~\ref{fig:UcVsAlphaaMode2} shows results for the instability waves of
mode $2$. Figure~\ref{fig:UcVsAlphaaMode2}(a) is for $N = 1$. We expect little
change caused by one single lobe, and this is demonstrated clearly by the
figure.
The results for $N = 2$ are shown in
figure~\ref{fig:UcVsAlphaaMode2}(b). The mode $2$ instability waves have
increasingly lower convection velocities when $\epsilon$ increases. But the
changes are very small. The changes of the temporal growth rate are nearly
unobservable. However, using $3$ lobes can still effectively reduce the
convection velocity while marginally altering the temporal growth rate of the
mode $2$ instability waves. Like all the results we have reported, the use
of $5$ lobes is the most effective way of increasing the convection velocity
and decreasing the temporal growth rate.
Figure~\ref{fig:UcVsAlphaaMode-2} shows the results for the mode $-2$
instability waves. Figure~\ref{fig:UcVsAlphaaMode-2}(a) exhibits expected
behaviour for a lobed profile of $N = 1$. Figure~\ref{fig:UcVsAlphaaMode-2}(b)
shows a very slight increase of the convection velocity and no change to the
temporal growth rate. We emphasize again that this is due to $\lambda_n \ne
\lambda_{-n}$. Figures~\ref{fig:UcVsAlphaaMode-2}(c) and
\ref{fig:UcVsAlphaaMode-2}(d) are identical to
figures~\ref{fig:UcVsAlphaaMode2}(c) and \ref{fig:UcVsAlphaaMode2}(d)
respectively because of identical eigenvalues. 

In summary, the stability characteristics of base flows of a lobed vortex-sheet
type are different from those of axisymmetric ones. The differences consist of
changes to both the convection velocity and the temporal growth rate of
instability waves. The changes become more pronounced as the number of lobes
$N$ and the penetration ratio $\epsilon$ increase. However, instability waves
with different mode numbers are affected differently by the lobed geometry. In
particular, little change occurs for mode $0$, no matter how large both $N$ and
$\epsilon$ are. On the other hand, an evident alteration of the characteristics
of jet instability waves with large mode numbers occurs when $N > 1$. For $N =
3$ and $N = 5$, azimuthally even and odd instability waves demonstrate the same
characteristics. However, for $N =2$ and $N = 1$, even and odd instability
waves of lobed jets exhibit two different types of behaviour, with one having
favourable effects on installed jet noise reduction and the other having
adverse. Therefore, for the sake of suppressing instability waves, or achieving
installed jet noise reduction, it is desired to use a lobed profile of large
$N$, such as $N = 5$, with a large penetration ratio. Because this results in a
larger reduction of the temporal growth rates, and hence may lead to weaker
instability waves.

Note that the results discussed above are obtained from a temporal analysis,
from which the results of a spatial analysis may be recovered from Gaster's
transformation~\citep{Gaster1962}. Gaster's transformation provides a
straightforward way to connect the temporal and spatial stability analyses. One
of the key assumptions used in Gaster's transformation is that both the spatial
and temporal growth rates have to be small, and under this assumption Gaster
showed that, to the first-order accuracy, the temporal angular frequency,
$\Re(\omega)$, rather than the convection velocity $\Re(\omega)/\alpha$, is the
same as the spatial angular frequency, and the temporal growth rate is equal to
the product of the spatial growth rate and the temporal group velocity. With
the temporal results available, we can therefore obtain the spatial growth rate
by using Gaster's transformation, provided the aforementioned assumption is not
violated. Care must be taken for the spatial convection velocity as the
$\Re(\omega)$ being identical does not imply the same thing for
$\Re(\omega)/\Re(\alpha)$ (see the spatial convection velocity of a round
vortex sheet given by \citet{Michalke1970} for example), in which case
high-order terms need to be incorporated to improve accuracy.

However, the aim of this study is to investigate the effects of lobes on the
growth rates of jet instability waves so as to examine the feasibility of
controlling installed jet noise using lobed jets, and these effects are
unlikely to be significantly different between the temporal or spatial
frameworks. For example, if the lobed geometry slightly reduces the temporal
growth rate, we would expect similar behavior for the spatial growth rate.
Therefore, in this paper, we focus on the temporal results only. Also note that
previous studies mainly focused on chevron jets, and we do not expect their
results to be identical to the stability results for the special type of lobed
base flow used in this study. 

\subsection{The change to mode shapes}
\begin{figure}
    \centering
    \includegraphics{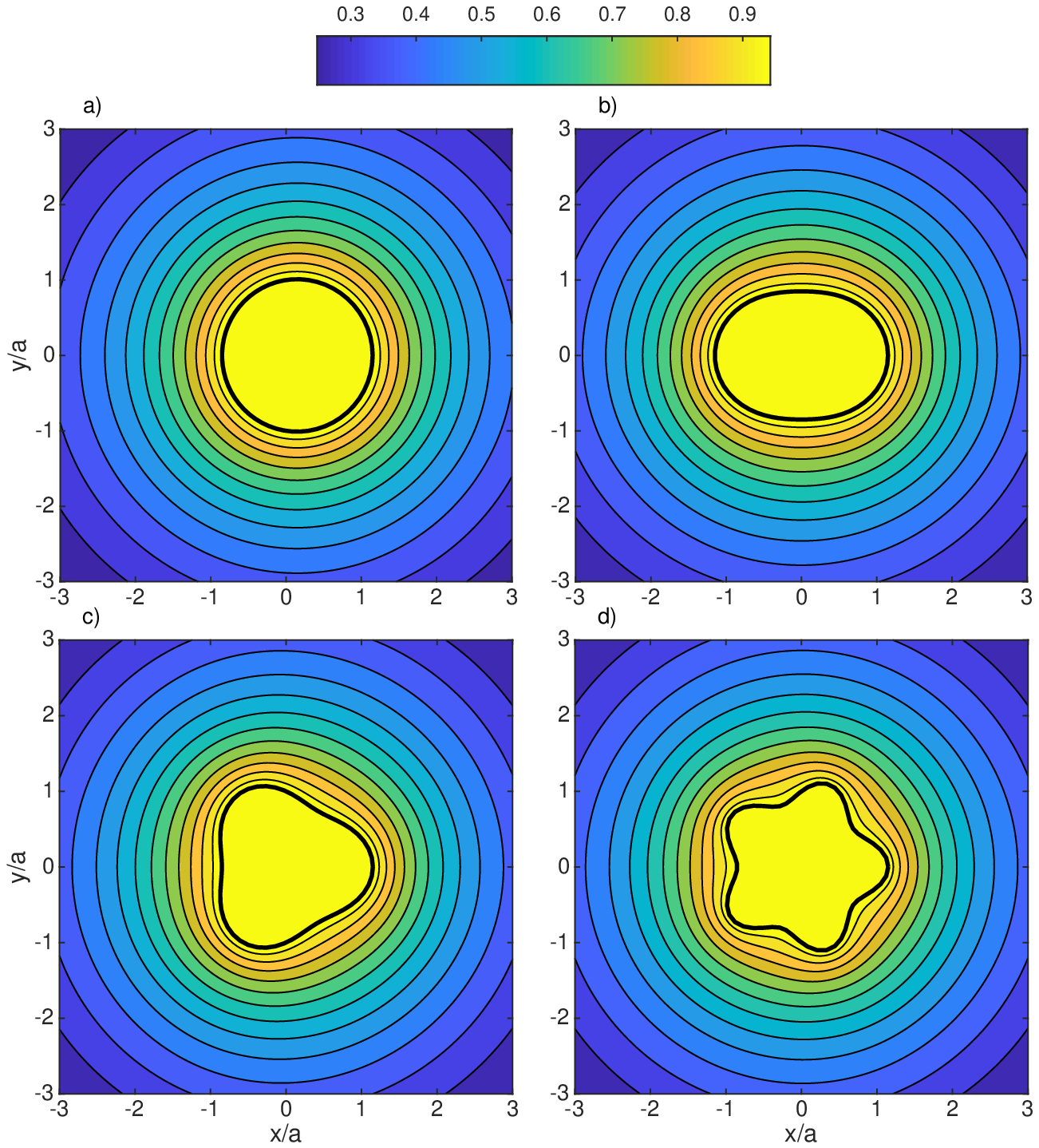}
    \caption{The shapes of mode 0 for some strongly lobed geometries when
    $\epsilon = 0.15$ at a low frequency of $\alpha a \approx  0.25$: a) $N =
    1$; b) $N = 2$; c) $N = 3$; d) $N = 5$. The thick black solid lines show
the lobed vortex sheets.}
    \label{fig:mode0LowFreq}
\end{figure}

\begin{figure}
    \centering
    \includegraphics{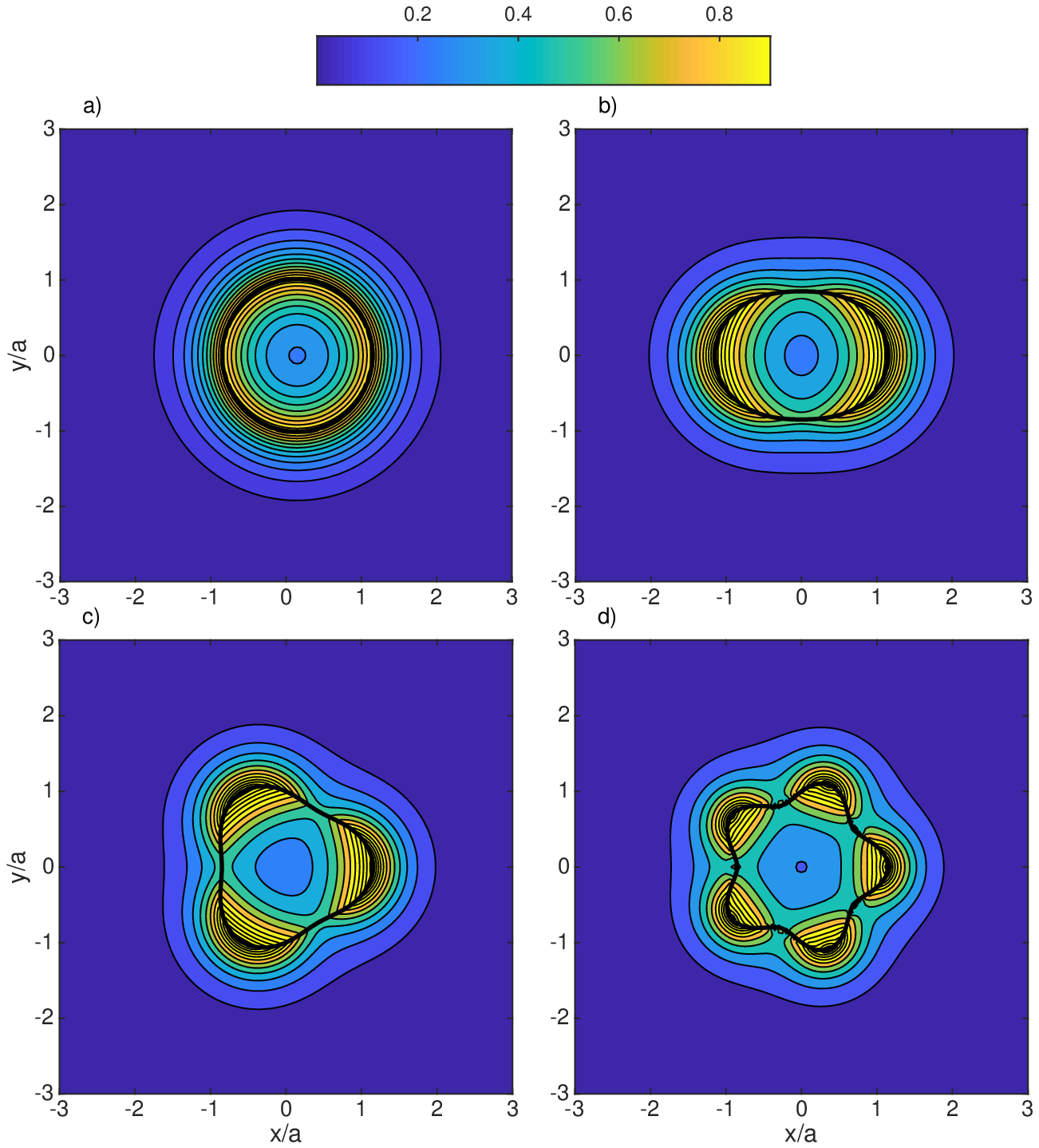}
    \caption{The shapes of mode 0 for some strongly lobed geometries when
    $\epsilon = 0.15$ at a high frequency of $\alpha a \approx 2.5$: a) $N =
    1$; b) $N = 2$; c) $N = 3$; d) $N = 5$. The thick black solid lines show
the lobed vortex sheets.}
    \label{fig:mode0HighFreq}
\end{figure}

\begin{figure}
    \centering
    \includegraphics{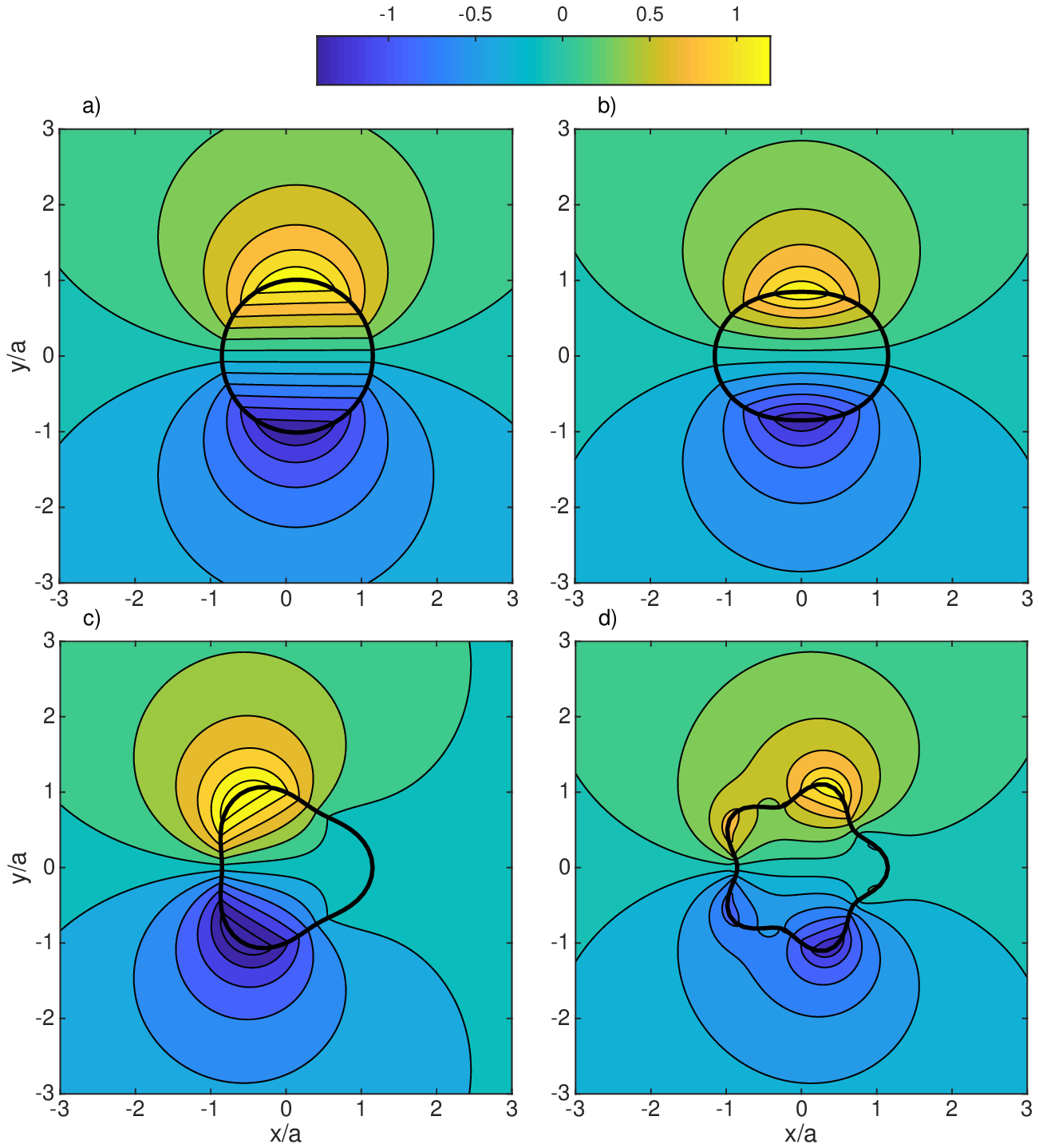}
    \caption{The shapes of mode -1 for some strongly lobed geometries when
    $\epsilon = 0.15$ at a low frequency of $\alpha a \approx  0.25$: a) $N =
    1$; b) $N = 2$; c) $N = 3$; d) $N = 5$. The thick black solid lines show
the lobed vortex sheets.}
    \label{fig:mode-1LowFreq}
\end{figure}

\begin{figure}
    \centering
    \includegraphics{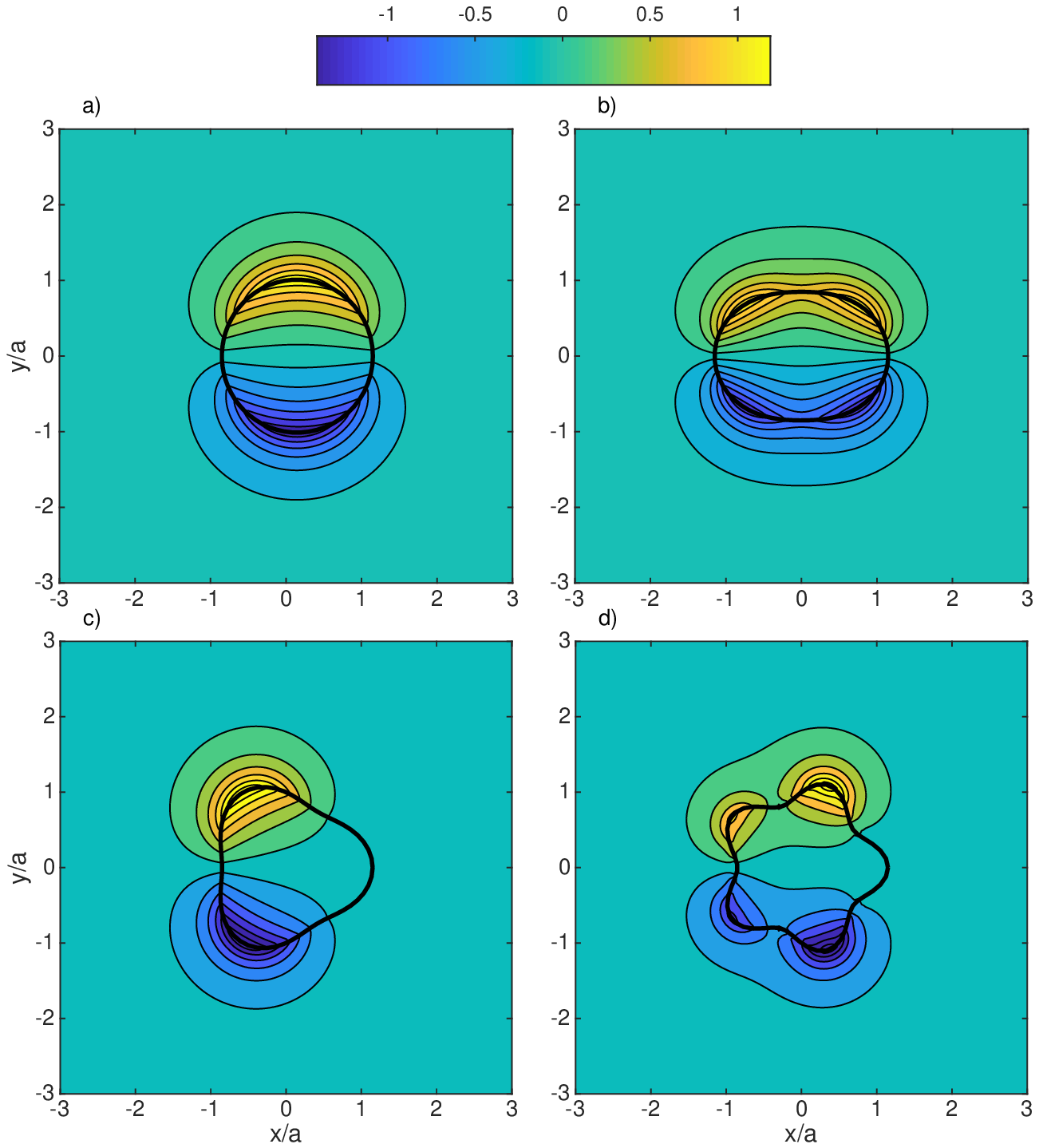}
    \caption{The shapes of mode -1 for some strongly lobed geometries when
    $\epsilon = 0.15$ at a high frequency of $\alpha a \approx 2.5$: a) $N =
    1$; b) $N = 2$; c) $N = 3$; d) $N = 5$. The thick black solid lines show
the lobed vortex sheets.}
    \label{fig:mode-1HighFreq}
\end{figure}

\begin{figure}
    \centering
    \includegraphics{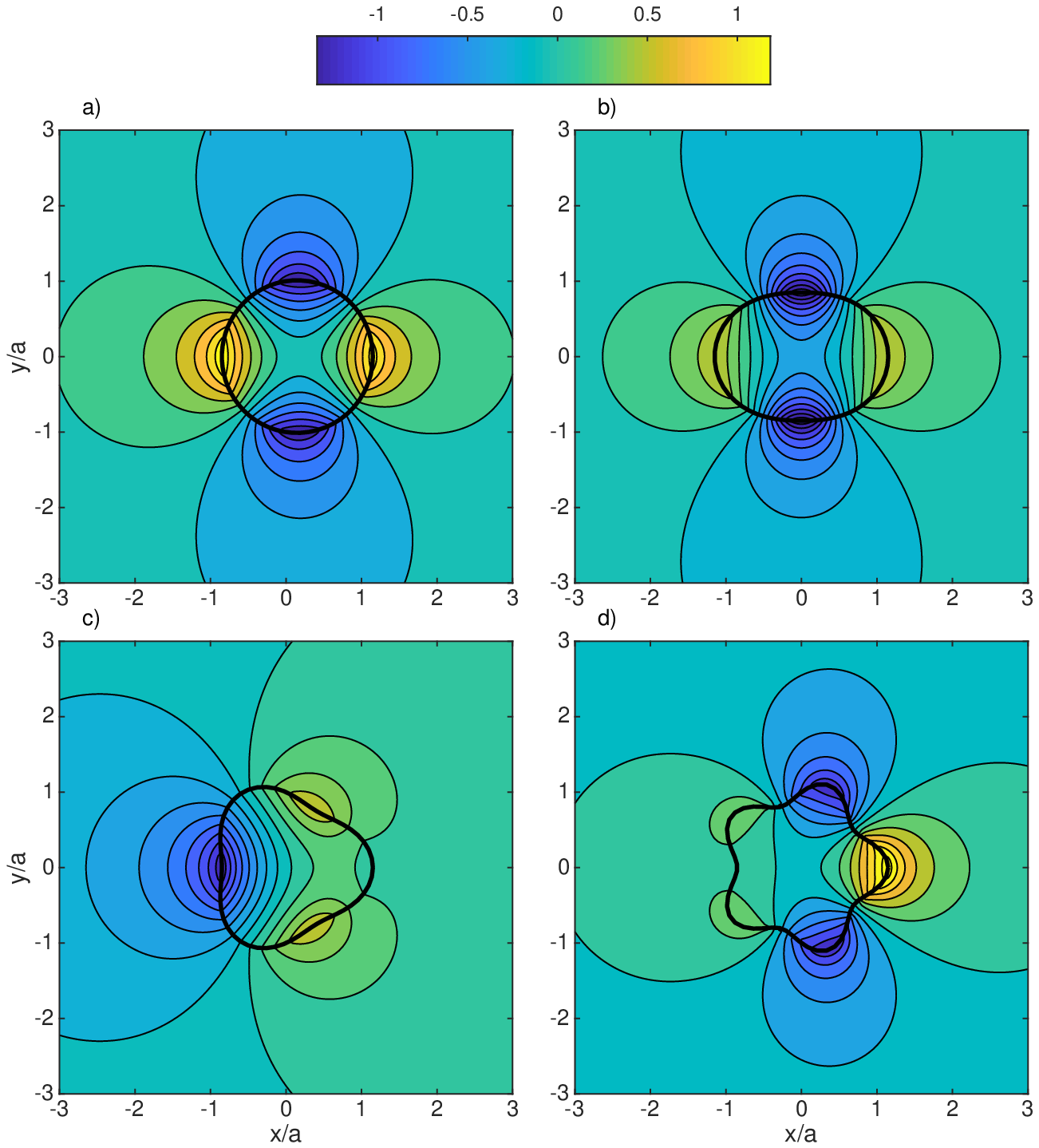}
    \caption{The shapes of mode 2 for some strongly lobed geometries when
    $\epsilon = 0.15$ at a low frequency of $\alpha a \approx  0.25$: a) $N =
    1$; b) $N = 2$; c) $N = 3$; d) $N = 5$. The thick black solid lines show
the lobed vortex sheets.}
    \label{fig:mode2LowFreq}
\end{figure}

\begin{figure}
    \centering
    \includegraphics{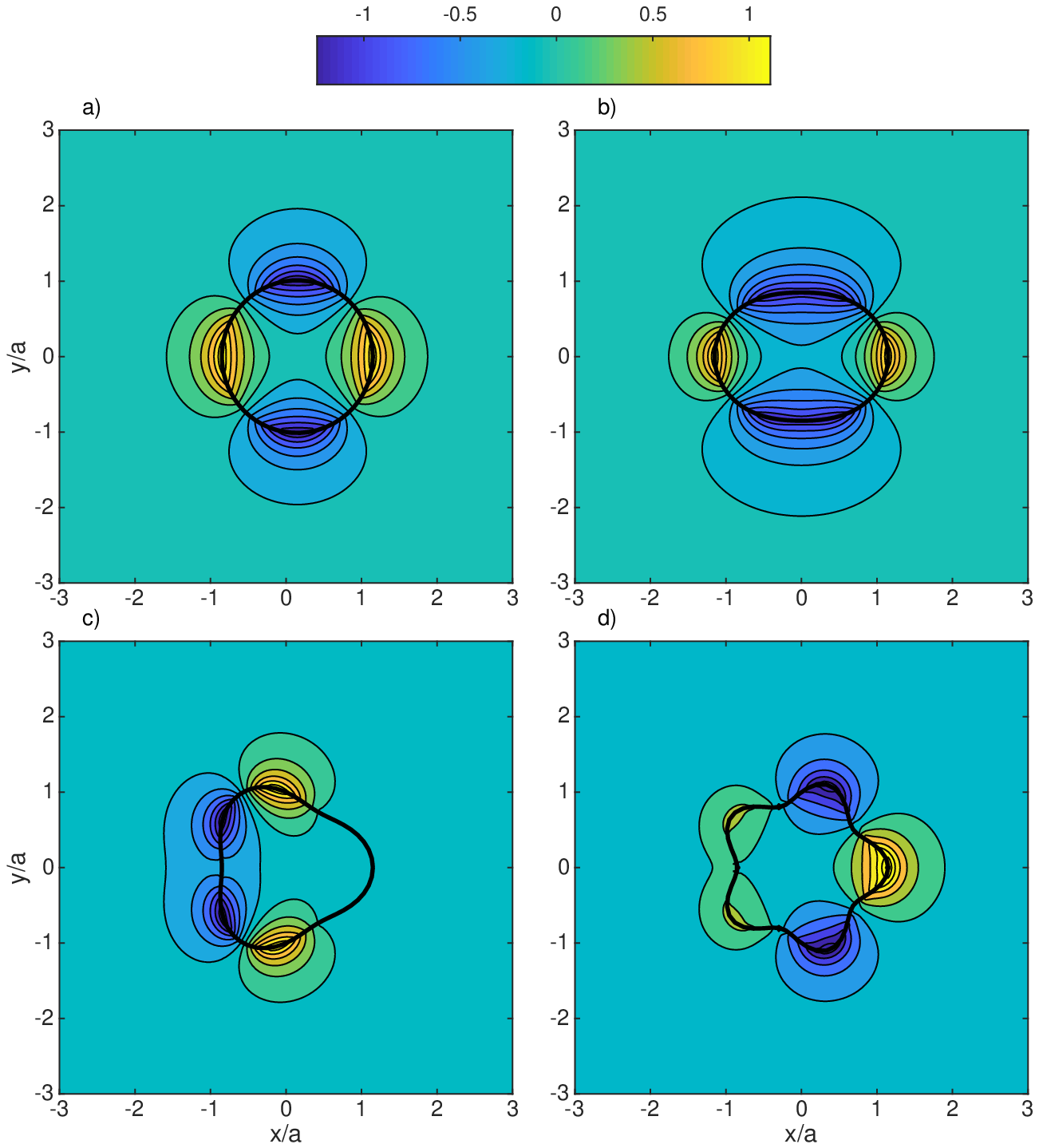}
    \caption{The shapes of mode 2 for some strongly lobed geometries when
    $\epsilon = 0.15$ at a high frequency of $\alpha a \approx 2.5$: a) $N =
    1$; b) $N = 2$; c) $N = 3$; d) $N = 5$. The thick black solid lines show
the lobed vortex sheets.}
    \label{fig:mode2HighFreq}
\end{figure}

We have studied the effects of lobed geometries on the convection velocity and
temporal growth rate of jet instability waves. In this section, we show their
effects on the mode shapes. In the rest of this section, we present the contour
plot of the pressure field corresponding to each eigen-mode. The pressure is
normalized such that it is equal to the right hand side of
equation~\ref{equ:dynamicBCUsingE} inside the vortex sheet. We choose to shown
the pressure fields for modes $0$, $-1$ and $2$, respectively. When doing so,
we fix $\epsilon = 0.15$ but let $N$ vary between $1$, $2$, $3$ and $5$,
respectively. For each mode at each value of $N$, we show the pressure
distribution at a low frequency $\alpha a \approx 0.25$ and then at a high
frequency $\alpha a \approx 2.5$. These results are shown from
figures~\ref{fig:mode0LowFreq} to \ref{fig:mode2HighFreq}. In each figure, the
lobed profile is represented by a thick solid black line.

Figure~\ref{fig:mode0LowFreq} shows the pressure distributions for mode $0$ at
$\alpha a \approx 0.25$. Figure~\ref{fig:mode0LowFreq}(a) is for $N = 1$. We
have mentioned that when $N = 1$ the lobed profile is more or less the same as
a displaced circle, and hence its stability characteristics should be nearly
the same as a round jet. Figure~\ref{fig:mode0LowFreq}(a) proves this by
showing a nearly axisymmetric pressure distribution. The pressure inside the
vortex sheet is nearly uniform. This is because the frequency is very low and
the modified Bessel function of the first kind approaches to a constant value
for a small argument. The pressure outside gradually decays to zero as the
distance to the centre of the vortex sheet increases. Note how the pressure is
matched continuously across the vortex sheet --- another indication for a
well-converged eigen-solution. Figure~\ref{fig:mode0LowFreq}(b) shows the
result for $N =2$. In this case the lobed profile resembles an ellipse. The
effect of the geometry change is to stretch the pressure field inside the
vortex sheet to have a distribution similar to the shape of the vortex sheet
itself. The pressure outside gradually becomes axisymmetric and decays to zero
at a large distance. The behaviours for $N = 3$ and $5$, as shown in
figures~\ref{fig:mode0LowFreq}(c) and \ref{fig:mode0LowFreq}(d), respectively,
are very similar to $N =2$.

Figure~\ref{fig:mode0HighFreq} shows the pressure fields for mode $0$ at a high
frequency $\alpha a \approx 2.5$. A striking difference is the much smaller
size of contour regions in each sub-figure. This is due to the fact that the
pressure decays more quickly outside the vortex sheet at high frequencies. This
is in accord with the fact that installed jet noise is only significant at low
frequencies. Figure~\ref{fig:mode0HighFreq}(a) again shows the result for $N =
1$. One difference from figure~\ref{fig:mode0LowFreq}(a) is that the pressure
variation inside the vortex sheet is clear at this high frequency, which is
what we would expect.  When $N =2$, the lobed geometry has the same stretching
effects as those shown in figure~\ref{fig:mode0LowFreq}(b), but with a more
marked pressure variation. The same tendencies can be seen from
figures~\ref{fig:mode0LowFreq}(c) and \ref{fig:mode0LowFreq}(d). The relatively
insignificant change to the shape of the mode $0$ instability wave, as shown in
figures~\ref{fig:mode0LowFreq} and \ref{fig:mode0HighFreq}, is consistent with
the fact that both the convection velocity and temporal growth rate remain
roughly the same as those for an axisymmetric jet.

Figures~\ref{fig:mode-1LowFreq} shows the results for mode $-1$ at $\alpha a
\approx 0.25$. From section~\ref{subsec:labellingStrategy}, we know that for
negative mode numbers the pressure distribution is antisymmetric with respect
to $\phi = 0$. This is reflected in figure~\ref{fig:mode-1LowFreq}.
Figure~\ref{fig:mode-1LowFreq}(a) very much resembles to that of an
axisymmetric jet. When $N = 2$, the pressure field still largely resembles to
figure~\ref{fig:mode-1LowFreq}(a). A clear change in mode shape occurs when $N
=3$, as shown in figure~\ref{fig:mode-1LowFreq}(c). We can see that a
three-lobe profile causes the two lobes of the pressure field to tilt towards
two lobes of the vortex sheet. This signifies a potentially large change in the
convection velocity and temporal growth rate. The mode shape at $N = 5$  shows
similar characteristics to that at $N = 3$. Moreover, there are some small and
locale changes in response to the appearance of local geometry corrugations. We
like to point out, however, that although changes are observed, the mode shapes
are still bearing the signature of conventional $\sin\phi$ behaviour,
suggesting the appropriateness of our earlier mode labelling strategy.

Figure~\ref{fig:mode-1HighFreq} is similar to figure~\ref{fig:mode-1LowFreq} in
almost every aspect, apart from the smaller size of the contour regions due to
the quicker decay of the instability waves outside the vortex sheet at high
frequencies. We therefore omit a repetitive description. Instead, we try to
understand the behaviour of non-identical eigenvalues, as discussed in
section~\ref{subsec:effects}, from the perspective of mode shapes.  

We have observed that for $N = 1$ and $N = 2$, the eigenvalue $\lambda_{-1}$ is
not degenerate any more, and consequently, only standing modes are allowed,
whereas travelling modes (in the azimuthal direction, similar to a behaviour of
$e^{\ri (m\phi - \omega t)}$) are allowed for $N = 3$ and $5$. To understand
the difference we can compare figures~\ref{fig:mode-1HighFreq}(b) and
\ref{fig:mode-1HighFreq}(c). We conclude that, if an eigenvalue is not
degenerate, then each of the symmetric  plane of the lobe profile must also be
a symmetric (or antisymmetric) plane of the eigen-mode. For example, in
figure~\ref{fig:mode-1HighFreq}(b), the lobed profile has two symmetric planes,
namely the horizontal and vertical planes, each of which is also a symmetric or
antisymmetric plane of the pressure field. This is consistent with the fact
that $\lambda_1$ is not degenerate. On the other hand, in
figure~\ref{fig:mode-1HighFreq}(c), the lobed profile has three symmetric
planes, but only one of them is the antisymmetric plane of the pressure field.
Hence, $\lambda_{-1}$ must be degenerate. Figure~\ref{fig:mode-1HighFreq}(d) is
very similar to figure~\ref{fig:mode-1HighFreq}(c). For an axisymmetric jet,
the vortex sheet profile has infinitely many symmetric planes, however only two
of them are the symmetric and antisymmetric planes of mode $-1$ pressure field.
Therefore $\lambda_{-1}$ is degenerate for a round jet.  Note that the vortex
sheet profile in figure~\ref{fig:mode-1HighFreq}(a) is close to a displaced
circle, but is not strictly one. It has a slight eccentricity and therefore has
a non-degenerate eigenvalue. 

The connection between the degeneracy and the mode shape can be understood as
follows. Take figure~\ref{fig:mode-1HighFreq}(c) as an example. We mentioned
that only one of three symmetric planes, which are angled by $120^\circ$ from
each other, is the antisymmetric plane of the pressure field. Now we can rotate
the pressure field by $120^\circ$, then the resulting pressure field must also
be an eigen-solution of the stability problem, with the same eigenvalue.
Therefore, the corresponding eigenvalue must be degenerate. The combination of
the eigenvectors corresponding to the same eigenvalue creates infinitely many
eigen-solutions, and two of them are those obtained by rotating
figure~\ref{fig:mode-1HighFreq}(c) by $120^\circ$ and $240^\circ$,
respectively.

We expect that the non-degeneracy is likely to occur when the mode number
$n$ is equal to $\pm N/2$, $\pm N$ etc. This is because, for mode $n$, the
pressure field normally has $2|n|$ lobes, and the symmetric and antisymmetric
planes of such a mode can be aligned with all the symmetric planes of the
vortex sheet profile. Comparing with the figures shown in
section~\ref{subsec:effects}, we find that this is indeed the case. For
example, we see $\lambda_{-1} \ne \lambda_1$ when $N = 2$ and $\lambda_{-2} \ne
\lambda_2$ when $N = 2$.

Figure~\ref{fig:mode2LowFreq} shows the distributions of the pressure field for
mode $2$ at the low frequency $\alpha a \approx 0.25$.
Figure~\ref{fig:mode2LowFreq}(a) shows its similarity to that for an
axisymmetric vortex sheet while figure~\ref{fig:mode2LowFreq}(b) for an
elliptic vortex sheet. Note how the earlier conclusion about non-degeneracy
remains valid in figure~\ref{fig:mode2LowFreq}(b-d).
Figure~\ref{fig:mode2HighFreq} shows qualitatively similar results at a
higher frequency so we avoid an unnecessary repetition.

\section{Conclusion}
In the hope of suppressing installed jet noise, an analytical study of the
stability characteristics of lobed jets of a vortex sheet type is performed. It
is shown that the lobed geometry changes both the convection velocity and the
temporal growth rate of the instability waves. The effects are more pronounced
as the number of lobes $N$ and the penetration ratio $\epsilon$ increase.
However, instability waves of different mode numbers are affected differently
by the lobes. For instance, the mode $0$ is particularly insensitive to the
geometry changes. Higher modes are more likely to be changed significantly when
both $N$ and $\epsilon$ are sufficiently large. An interesting finding is that
when $N=1$ and $N=2$ different behaviour occurs between the even and odd
instability waves of certain orders, i.e. the corresponding eigenvalue becomes
non-degenerate. We show that a necessary condition for a non-degenerate
eigenvalue to exist is that each symmetric plane of the lobed vortex sheet must
also be a symmetric or antisymmetric plane of the corresponding mode shape.
This is likely to occur when the mode number $n$ is $\pm N/2$, $\pm N$ and so
on. It is concluded that in order to suppress instability waves for the sake of
reducing installed jet noise, a large $N$, such as $N=5$, and a large
$\epsilon$ are desirable. 

The insensitiveness of the mode $0$ instability waves to the lobed geometry
implies that using lobed geometry hardly helps in reducing the installed jet
noise due to the scattering of the mode $0$ jet instability waves. The
reduction of the total installed jet noise would be somewhat limited. If the
modes $0$ and $1$ (including both modes $+1$ and $-1$) instability waves are of
equal strength, one would expect an observable sound reduction up to $3$ dB (in
the ideal case). It is, however, worth noting that the current analysis is
based on a parallel vortex sheet assumption. In realistic jets, the jet mean
flow is expected to gradually become axisymmetric downstream the jet exit due
to strong jet mixing. To what extent this smoothing of the lobed geometry would
affect the jet stability, hence installed jet noise, requires further
examination. This constitutes part of our future work. 
\section*{Acknowledgement}
The first author (B. L.) wishes to gratefully acknowledge the financial support
provided by the Cambridge Commonwealth European and International Trust and the
China Scholarship Council.

\appendix
\section{The kinematic and dynamic boundary conditions}
\label{app:BC}
The two equations shown in equation~\ref{equ:expansion3}, together with the
kinematic and dynamic boundary conditions on the vortex sheet, need to be
combined to obtain the dispersion relation. The kinematic and dynamic
boundary conditions can be obtained as follows. As defined above, the vortex
sheet profile is given by $\mathcal{F}(\sigma, \phi) = \sigma -
\mathcal{R}(\phi) = 0$. One can assume that the perturbed profile can be
described by the function
\begin{equation}
    \mathcal{F}_p(\sigma, \phi, z, t) = \mathcal{F}(\sigma, \phi) -
    \eta^\prime(\phi, z, t) = 0,   
\end{equation}
where $\eta^\prime(\phi, z, t)$ denotes a small-amplitude perturbation of the
radius of the vortex-sheet profile. The kinematic boundary condition states
that, on the perturbed vortex sheet,
\begin{equation}
    \frac{\D \mathcal{F}_p(\sigma, \phi, z, t)}{\D t} = 0.
    \label{equ:dynamicBoundaryConditionLobed}
\end{equation}
Substituting the velocity on both sides of the vortex sheet to
equation~\ref{equ:dynamicBoundaryConditionLobed} and linearising around
the unperturbed vortex sheet yields
\begin{equation}
    \begin{dcases}
	\frac{\partial \eta^\prime}{\partial t} + U \frac{\partial
	\eta^\prime}{\partial z} 
	- \nabla{\psi}^- \cdot \nabla \mathcal{F} = 0,\\
	\frac{\partial \eta^\prime}{\partial t} 
	- \nabla {\psi}^+ \cdot \nabla \mathcal{F} = 0.
    \end{dcases}
    \label{equ:twoDynamicBoundaryConditions}
\end{equation}
After invoking the harmonic time and $z$ dependence ($\e^{\ri \alpha z}
\e^{-\ri \omega t}$) and eliminating $\eta^\prime$, one can
show that equation~\ref{equ:twoDynamicBoundaryConditions} reduces to 
\begin{equation}
    (\omega - \alpha U) \nabla \psi^{+} \cdot \boldsymbol{n} = \omega
    \nabla \psi^{-} \cdot \boldsymbol{n},
\end{equation}
where $\boldsymbol{n}$ denotes the unit vector perpendicular to the
vortex-sheet profile, which can be readily shown to be $\nabla
\mathcal{F}/|\nabla \mathcal{F}|$. The dynamic boundary condition requires
pressure continuity across the vortex sheet. From the linearized momentum
equation, i.e. equation~\ref{equ:linearisedMomentumEquationInvscid}, one can
readily show that 
\begin{equation}
    p^\prime = - \rho \left(\frac{\partial }{\partial t} + U_z
    \frac{\partial }{\partial z}\right) \psi,
\end{equation}
where $U_z$ can be either $U$ and $0$ depending on which side of the vortex
sheet is considered. Hence substituting the velocity potentials on both sides
of the vortex sheet yields that on the unperturbed vortex sheet (after
linearising around the unperturbed vortex sheet)
\begin{equation}
    \omega \psi^{+} = (\omega - \alpha U) \psi^{-},
\end{equation}

\nomenclature[Rf]{$\mathcal{F}_p$}{Function defining the perturbed vortex
sheet profile}
\nomenclature[Rn]{$\boldsymbol{n}$}{Unit vector perpendicular to the lobed
profile}
\nomenclature[Ru]{$U_z$}{Defined to be $U$ inside but $0$ outside of the vortex sheet}

\section{The properties on rotationally symmetric matrices}
\label{app:rotationalsymmetry}
Since the eigenvector $\boldsymbol{C}^-$ of the matrix $\boldsymbol{A}$ fully
determines the eigenfunction $E_n^-(\sigma, \phi)$, it is important to examine
the properties of $\boldsymbol{A}$ and its eigenvectors. In order to do this,
we need to define the rotation of any vector $\boldsymbol{C}$ as
$\boldsymbol{C}^R$, such that
\begin{equation}
    (\boldsymbol{C}^R)_n = C_{-n},
    \label{equ:reverseVector}
\end{equation}
where $(\boldsymbol{C}^R)_n$ denotes the $n$th the element of the
rotation vector $\boldsymbol{C}^R$ and $C_n$ is the $n$th element of
$\boldsymbol{C}$. Note here the index of the vector $\boldsymbol{C}$ ranges
from $-M$ to $M$. Similarly, we can define the rotation of any matrix
$\boldsymbol{B}$ as
\begin{equation}
    (\boldsymbol{B}^R)_{ij} =  B_{(-i)(-j)},
    \label{equ:reverserMatrix}
\end{equation}
where $(\boldsymbol{B}^R)_{ij}$ denotes the element of
$\boldsymbol{B}^R$ at the $i$th row and the $j$th column, and $B_{ij}$
are the indexed elements of matrix $\boldsymbol{B}$. If a vector
$\boldsymbol{C}$ is equal to its rotation, we define it as symmetric. If, on
the other hand, $\boldsymbol{C} = -\boldsymbol{C}^R$, we define it as
antisymmetric. Similarly, if a matrix $\boldsymbol{B}$ is equal to its
rotation, we define it as rotationally symmetric. Anti-rotational symmetry
follows a self-explanatory definition. One can now show that $(\boldsymbol{BC})^R
= \boldsymbol{B}^R \boldsymbol{C}^R$, because
\begin{equation}
    \big((\boldsymbol{B}\boldsymbol{C})^R\big)_i 
    = \sum_{j = -M}^M B_{(-i)j}C_j  = \sum_{j = -M}^M B_{(-i)(-j)}C_{-j} =
    \big(\boldsymbol{B}^R \boldsymbol{C}^R\big)_i,
\end{equation}
where $i$ can be any number between $-M$ to $M$. Replacing the vector
$\boldsymbol{C}$ in the above equations with a matrix $\boldsymbol{B}_2$ does
not invalidate the formula, i.e. $(\boldsymbol{BB}_2)^R
=\boldsymbol{B}^R \boldsymbol{B}_2^R$ also holds. Therefore, if
both $\boldsymbol{B}$ and $\boldsymbol{B}_2$ are rationally symmetric
matrices, then the product of them is also rotationally symmetric. This is
because
\begin{equation}
    (\boldsymbol{BB}_2)^R = \boldsymbol{B}^R
    \boldsymbol{B}_2^R  = \boldsymbol{BB}_2.
\end{equation}

\nomenclature[Rc]{$\boldsymbol{C}$}{General vector}
\nomenclature[Rb]{$\boldsymbol{B}$}{General matrix}
\nomenclature[Rb]{$\boldsymbol{B}_2$}{General matrix}

We now prove that the inverse of a rotationally symmetric matrix, if exists,
is also rotationally symmetric. First, because it is assumed that the inverse
of the rotationally symmetric $\boldsymbol{B}$ exists, we denote its column
vectors by $\boldsymbol{D}_i$ ($i = -M\dots M$), i.e. 
\begin{equation}
    \boldsymbol{B}^{-1} = [\boldsymbol{D}_{-M}\dots
    \boldsymbol{D}_0\dots\boldsymbol{D}_M].
\end{equation}
Then according to the definition of the inverse matrix, one has 
\begin{equation}
    [\boldsymbol{B}\boldsymbol{D}_{-M}\dots
    \boldsymbol{B}\boldsymbol{D}_0\dots\boldsymbol{B}\boldsymbol{D}_M] =
    [\boldsymbol{I}_{-M}\dots \boldsymbol{I}_0\dots\boldsymbol{I}_M],
    \label{equ:submatrices}
\end{equation}
where $\boldsymbol{I}_i$ is the $i$th column of the identity matrix. Now for
any positive number $i$, one has
\begin{equation}
    \boldsymbol{B}\boldsymbol{D}_i = \boldsymbol{I}_i.
    \label{equ:columwiseEquation}
\end{equation}
If taking the rotation of both sides of
equation~\ref{equ:columwiseEquation}, one obtains
\begin{equation}
    \boldsymbol{B} \boldsymbol{D}_i^R = \boldsymbol{I}_{-i},
    \label{equ:daggerEquation}
\end{equation}
where use is made of the fact that $\boldsymbol{I}_{i}^R =
\boldsymbol{I}_{-i}$ and $\boldsymbol{B}$ is equal to its own rotation.
Comparing equation~\ref{equ:daggerEquation} with the $-i$ column of
equation~\ref{equ:submatrices}, we have $\boldsymbol{D}_i^R =
\boldsymbol{D}_{-i}$. This is because $\boldsymbol{B}$ is invertible, its has a
full rank and the solution of equation~\ref{equ:daggerEquation} is unique. We
have now proved that $\boldsymbol{B}^-$ is indeed a rationally symmetric
matrix.

\nomenclature[Rd]{$\boldsymbol{D}_i$}{Column vectors of matrix $\boldsymbol{B}$
($i=\dots -2, 1,2 \dots$)}
\nomenclature[Ri]{$\boldsymbol{I}_i$}{Column vectors of the identity matrix
($i=\dots -2, 1,2 \dots$)} 

Examining the definitions of all the relevant $\boldsymbol{I}$ and
$\boldsymbol{K}$ matrices, it is trivial to show that they are all
rotationally symmetric. Based on the two conclusions discussed above, because
the matrix $\boldsymbol{A}$ can be written as
\begin{equation*}
    \boldsymbol{A} = \widetilde{\boldsymbol{I}}_k^{-1}
    \widetilde{\boldsymbol{K}}_k \widetilde{\boldsymbol{K}}_d^{-1}
    \widetilde{\boldsymbol{I}}_d, 
\end{equation*}
it can be easily shown that $\boldsymbol{A}$ is rotationally symmetric. One
important property that follows is that if a vector $\boldsymbol{C}^-$ is one
of the eigenvectors of $\boldsymbol{A}$, so is $\boldsymbol{C}^{-R}$.
This follows naturally after taking the rotation of both sides of the
eigenvalue equation $\boldsymbol{A} \boldsymbol{C}^- = \lambda
\boldsymbol{C}^-$. One consequence of this property is that if an eigenvalue
$\lambda_n$ of matrix $\boldsymbol{A}$ has no multiplicity, its eigenvector
$\boldsymbol{C}^-$ must be either symmetric or antisymmetric. The second
important property is that, for each multiple-folded eigenvalue $\lambda_n$, we
can always construct both a symmetric (e.g. $\boldsymbol{C}^- +
\boldsymbol{C}^{-R}$) and an antisymmetric (e.g. $\boldsymbol{C}^- -
\boldsymbol{C}^{-R}$) eigenvector. These properties are essential when we
try to assign an order to each obtained eigenvector in
Appendix~\ref{app:labellingStrategy}.

\section{The mode labelling strategy}
\label{app:labellingStrategy}
It is not difficult to show that, when $\epsilon = 0$, $\boldsymbol{A}$ is
diagonal and its eigenvalues (diagonal elements) are 
\begin{equation}
    \lambda_n = \frac{K_n^\prime(\alpha a) I_n(\alpha a)}{K_n(\alpha a)
    I_n^\prime(\alpha a)}
    \label{equ:dispersionRelationshipCylindricalVortexFromMatrix}
\end{equation}
and their corresponding normalized eigenvectors are
\begin{equation}
    \boldsymbol{C}^- = \left[ \hdots, 0, \hdots, 0, \hdots,
    1, \hdots \right]^T,
    \label{equ:eigenvectorRoundJet}
\end{equation}
where $1$ appears at the position of $C_n^-$. The well-known results for the
cylindrical vortex-sheet flow are recovered. When $\epsilon$ increases
gradually, we expect that the eigenvector gradually changes to 
\begin{equation}
    \boldsymbol{C}^- = \left[ \hdots, a_{-n}, \hdots, a_{0}, \hdots,
    1-a_{n}, \hdots \right]^T,
    \label{equ:eigenvectorLobedJet}
\end{equation}
where $a_n$ are complex numbers and $|a_n| \ll 1$. We may use
this dominant-component property of $\boldsymbol{C}^-$ to label the order of
the eigenfunctions. That is, the eigenfunction determined by the eigenvector 
\begin{equation}
    \boldsymbol{C}^- = \left[ \hdots, C_{-n}^-, \hdots, C_{0}^-, \hdots, C_{n}^-,
    \hdots \right]^T
\end{equation}
has an mode number $n$, if
\begin{equation}
    ||\boldsymbol{C}^-  - \boldsymbol{G}|| = \sqrt{\sum_{j=-M}^M
    (|C_j^-| - G_j)^2}
    \label{equ:travellingEigenvector}
\end{equation}
yields a minimum value when 
\begin{equation}
    \boldsymbol{G} = \left[ \hdots, 0, \hdots, 0, \hdots,
    g_n = 1, \hdots \right]^T,
\end{equation}
where $g_n$ is the element of the vector $\boldsymbol{G}$.

\nomenclature[Ra]{$a_i$}{Complex numbers of small moduli}
\nomenclature[Rg]{$\boldsymbol{G}$}{Gauge vector}
\nomenclature[Rg]{$g_n$}{Components of Gauge vector $\boldsymbol{G}$}

This strategy, however, hinges on the assumption that there is only
one dominant component in the eigenvector $\boldsymbol{C}^-$. However, this is
not always possible. For example, when $\epsilon \ne 0$, $\lambda_n$ and
$\lambda_{-n}$ do not necessarily have to be same any more. As we discussed
above, the eigenvector must be either symmetric or antisymmetric. Since both
symmetric and antisymmetric eigenvectors have two dominant components at $n$
and $-n$ respectively ($\lambda_n \ne \lambda_{-n}$, therefore one cannot
perform a linear combination of the two corresponding eigenvectors), it is hard
to determine whether this mode should be called mode $n$ or $-n$. 

To overcome this problem, we force each eigenvector to be either symmetric or
antisymmetric. This is always possible, because, as we proved earlier, the
eigenvector $\boldsymbol{C}^-$ must be either symmetric or antisymmetric if
$\lambda_n \ne \lambda_{-n}$. And if $\lambda_n = \lambda_{-n} (n>0)$, we can
always make use the second property of the eigenvectors and redefine one of the
two corresponding eigenvectors to be symmetric and the other antisymmetric. In
doing so, no matter whether $\lambda_n$ and $\lambda_{-n}$ are equal or not,
each eigenvector would have two dominant components. Now for the symmetric
eignvectors if $||\boldsymbol{C}^- - \boldsymbol{G}||$ obtains its minimum when 
\begin{equation}
    \boldsymbol{G} = \left[ \hdots, g_{-n} = \sqrt{2}/2, \hdots, 0, \hdots,
    g_n = \sqrt{2}/2, \hdots \right]^T,
    \label{equ:symmetricEigenvectorRepeated}
\end{equation}
we label the eigenvector as mode $n$. For anti-symmetric eigenvectors we label
it as $-n$ in a similar manner. For $n=0$, it is trivial to label its mode
number, and because we require $C_0^- \ne 0$, it can be shown that the
eigenvector must be symmetric. By labelling the eigenvectors in this way, from
equation~\ref{equ:EnExpansion}, this means that all nonnegative eigenfunctions
are even functions of $\phi$ and negative ones odd.

\bibliography{cleanRef}

\begin{thebibliography}{25}
\expandafter\ifx\csname natexlab\endcsname\relax\def\natexlab#1{#1}\fi
\def\au#1{#1} \def\ed#1{#1} \def\yr#1{#1}\def\at#1{#1}\def\jt#1{\textit{#1}}
  \def\bt#1{#1}\def\bvol#1{\textbf{#1}} \def\vol#1{#1} \def\pg#1{#1}
  \def\publ#1{#1}\def\arxiv#1{#1}\def\org#1{#1}\def\st#1{\textit{#1}}

\bibitem[Batchelor \& Gill(1962)]{Batchelor1962}
{\sc \au{Batchelor, G.~K.} \& \au{Gill, A.~E.}} \yr{1962}  \at{Analysis of the
  stability of axisymmetric jets}.  \jt{Journal of Fluid Mechanics}  \bvol{14},
   \pg{529--551}.

\bibitem[Baty \& Morris(1995)]{Baty1995}
{\sc \au{Baty, R.~S.} \& \au{Morris, P.~J.}} \yr{1995}  \at{The instability of
  jets of arbitrary exit geometry}.  \jt{International Journal for Numerical
  Methods in Fluids}  \bvol{21}~(9),  \pg{763--780}.

\bibitem[Cavalieri {\em et~al.\/}(2014)Cavalieri, Jordan, Wolf \&
  Gervais]{Cavalieri2014}
{\sc \au{Cavalieri, A. V.~G.}, \au{Jordan, P.}, \au{Wolf, W.} \& \au{Gervais,
  Y.}} \yr{2014}  \at{Scattering of wavepackets by a flat plate in the vicinity
  of a turbulent jet}.  \jt{Journal of Sound and Vibration}  \bvol{333},
  \pg{6516--6531}.

\bibitem[Cavalieri {\em et~al.\/}(2013)Cavalieri, Rodr\'{i}guez, Jordan,
  Colonius \& Gervais]{Cavalieri2013}
{\sc \au{Cavalieri, A. V.~G.}, \au{Rodr\'{i}guez, D.}, \au{Jordan, P.},
  \au{Colonius, T.} \& \au{Gervais, Y.}} \yr{2013}  \at{Wavepackets in the
  velocity field of turbulent jets}.  \jt{Journal of Fluid Mechanics}
  \bvol{730},  \pg{559--592}.

\bibitem[Crighton(1973)]{Crighton1973}
{\sc \au{Crighton, D.~G.}} \yr{1973}  \at{Instability of an elliptic jet}.
  \jt{Journal of Fluid Mechanics}  \bvol{59}~(4),  \pg{665--672}.

\bibitem[Gaster(1962)]{Gaster1962}
{\sc \au{Gaster, M.}} \yr{1962}  \at{A note on the relation between
  temporally-increasing and spatially-increasing disturbances in hydrodynamic
  stability}.  \jt{Journal of Fluid Mechanics}  \bvol{14},  \pg{222--224}.

\bibitem[Hu {\em et~al.\/}(2002)Hu, Saga, Kobayashi \& Taniguchi]{Hu2002}
{\sc \au{Hu, H.}, \au{Saga, T.}, \au{Kobayashi, T.} \& \au{Taniguchi, N.}}
  \yr{2002}  \at{Mixing process in a lobed jet flow}.  \jt{AIAA Journal}
  \bvol{40}~(7).

\bibitem[Kawahara {\em et~al.\/}(2003)Kawahara, Jim{\'e}nez, Uhlmann \&
  Pinelli]{Kawahara2003}
{\sc \au{Kawahara, G.}, \au{Jim{\'e}nez, J.}, \au{Uhlmann, M.} \& \au{Pinelli,
  A.}} \yr{2003}  \at{Linear instability of a corrugated vortex sheet - a model
  for streak instability}.  \jt{Journal of Fluid Mechanics}  \bvol{483},
  \pg{315--342}.

\bibitem[Kopiev {\em et~al.\/}(2004)Kopiev, Ostrikov, Chernyshev \&
  Elliot]{Kopiev2004}
{\sc \au{Kopiev, V.~F.}, \au{Ostrikov, N.~N.}, \au{Chernyshev, S.~A.} \&
  \au{Elliot, J.~W.}} \yr{2004}  \at{Aeroacoustics of supersonic jet issued
  from corrugated nozzle: new approach and prospects}.  \jt{International
  Journal of Aeroacoustics}  \bvol{3}~(3),  \pg{199--228}.

\bibitem[{Laj\'{u}s Jr.} {\em et~al.\/}(2015){Laj\'{u}s Jr.}, Cavalieri \&
  Deschamps]{Lajus2015}
{\sc \au{{Laj\'{u}s Jr.}, F.~C.}, \au{Cavalieri, A. V.~G.} \& \au{Deschamps,
  C.~J.}} \yr{2015} Spatial stability characteristics of non-circular jets.
  \bt{In {\em Proceedings of 21st AIAA/CEAS Aeroacoustics Conference\/}}.
  \publ{American Institute of Aeronautics and Astronautics}, 2015-2537.

\bibitem[Li {\em et~al.\/}(2001)Li, Hu, Kobayashi, Saga \& Taniguchi]{Li2001}
{\sc \au{Li, H.}, \au{Hu, H.}, \au{Kobayashi, T.}, \au{Saga, T.} \&
  \au{Taniguchi, N.}} \yr{2001}  \at{Visualization of multi-scale turbulent
  structure in lobed mixing jet using wavelets}.  \jt{Journal of Visulization}
  \bvol{4}~(3),  \pg{231--238}.

\bibitem[Li {\em et~al.\/}(2002)Li, Hu, Kobayashi, Saga \& Taniguchi]{Li2002}
{\sc \au{Li, Hui}, \au{Hu, Hui}, \au{Kobayashi, Toshio}, \au{Saga, Tetsuo} \&
  \au{Taniguchi, Nobuyuki}} \yr{2002}  \at{Wavelet multiresolution analysis of
  stereoscopic particle-image-velocimetry measurements in lobed jet}.  \jt{AIAA
  Journal}  \bvol{40}~(6).

\bibitem[Lyu {\em et~al.\/}(2017)Lyu, Dowling \& Naqavi]{Lyu2016d}
{\sc \au{Lyu, B.}, \au{Dowling, A.} \& \au{Naqavi, I.}} \yr{2017}
  \at{Prediction of installed jet noise}.  \jt{Journal of Fluid Mechanics}
  \bvol{811},  \pg{234--268}.

\bibitem[Lyu \& Dowling(2016)]{Lyu2016b}
{\sc \au{Lyu, B.} \& \au{Dowling, A.~P.}} \yr{2016} Noise prediction for
  installed jets.  \bt{In {\em Proceedings of 22nd AIAA/CEAS Aeroacoustics
  Conference\/}}.  \publ{American Institute of Aeronautics and Astronautics},
  aIAA 2016-2986.

\bibitem[Mankbadi \& Liu(1984)]{Mankbadi1984}
{\sc \au{Mankbadi, R.} \& \au{Liu, J. T.~C.}} \yr{1984}  \at{Sound generated
  aerodynamically revisited: large-scale structures in a turbulent jet as a
  source of sound}.  \jt{Philosophical Transactions of the Royal Society of
  London A}  \bvol{311}~(1516),  \pg{183--217}.

\bibitem[Miao {\em et~al.\/}(2015)Miao, Dmitriy, Liu, Qin, Wu, Chu \&
  Wang]{Miao2015}
{\sc \au{Miao, T.}, \au{Dmitriy, L.}, \au{Liu, J.}, \au{Qin, S.}, \au{Wu, D.},
  \au{Chu, N.} \& \au{Wang, L.}} \yr{2015}  \at{Study on the flow and acoustic
  characteristics of submerged exhaust through a lobed nozzle}.  \jt{Acoustics
  Australia}  \bvol{43},  \pg{283--293}.

\bibitem[Michalke(1970)]{Michalke1970}
{\sc \au{Michalke, A.}} \yr{1970}  \at{A wave model for sound generation in
  circular jets}.  \jt{Technical report} Deutsche Luftund Raumfahrt.

\bibitem[Morris(1988)]{Morris1988}
{\sc \au{Morris, P.~J.}} \yr{1988}  \at{Instability of elliptic jets}.
  \jt{AIAA Journal}  \bvol{26}~(2),  \pg{172--178}.

\bibitem[Morris(2010)]{Morris2010}
{\sc \au{Morris, P.~J.}} \yr{2010}  \at{The instability of high speed jets}.
  \jt{International Journal of Aeroacoustics}  \bvol{9}~(1-2),  \pg{1--50}.

\bibitem[Piantanida {\em et~al.\/}(2016)Piantanida, Jaunet, Huber, Wolf, Jordan
  \& Cavalieri]{Piantanida20164350}
{\sc \au{Piantanida, S.}, \au{Jaunet, V.}, \au{Huber, J.}, \au{Wolf, W.R.},
  \au{Jordan, P.} \& \au{Cavalieri, A.V.G.}} \yr{2016}  \at{Scattering of
  turbulent-jet wavepackets by a swept trailing edge}.  \jt{Journal of the
  Acoustical Society of America}  \bvol{140}~(6),  \pg{4350--4359}, cited By 2.

\bibitem[Sinha {\em et~al.\/}(2016)Sinha, Gudmundsson, Xia \&
  Colonius]{Sinha2016}
{\sc \au{Sinha, A.}, \au{Gudmundsson, K.}, \au{Xia, H.} \& \au{Colonius, T.}}
  \yr{2016}  \at{Parabolized stability analysis of jets from serrated nozzles}.
   \jt{Journal of Fluid Mechanics}  \bvol{789},  \pg{36--63}.

\bibitem[Tam \& Thies(1993)]{Tam1993}
{\sc \au{Tam, C. K.~W.} \& \au{Thies, A.~T.}} \yr{1993}  \at{Instability of
  rectangular jets}.  \jt{Journal of Fluid Mechanics}  \bvol{248},
  \pg{425--448}.

\bibitem[Tam \& Zaman(2000)]{Tam2000}
{\sc \au{Tam, C. K.~W.} \& \au{Zaman, K. B. M.~Q.}} \yr{2000}  \at{Subsonic jet
  noise from nonaxisymmetric and tabbed nozzles}.  \jt{AIAA Journal}
  \bvol{38}~(4),  \pg{592--599}.

\bibitem[Tinney \& Jordan(2008)]{Tinney2008b}
{\sc \au{Tinney, C.~E.} \& \au{Jordan, P.}} \yr{2008}  \at{The near pressure
  field of co-axial subsonic jets}.  \jt{Journal of Fluid Mechanics}
  \bvol{611},  \pg{175--204}.

\bibitem[Zaman {\em et~al.\/}(2003)Zaman, Wang \& Georgiadis]{Zaman2003}
{\sc \au{Zaman, K. B. M.~Q.}, \au{Wang, F.~Y.} \& \au{Georgiadis, N.~J.}}
  \yr{2003}  \at{Noise, turbulence and thrust of subsonic free jets from lobed
  nozzles}.  \jt{AIAA Journal}  \bvol{41}~(3),  \pg{389--407}.

\end{thebibliography}
\bibliographystyle{jfm}
\end{document}